\documentclass[useAMS,usenatbib]{mn2e}

\usepackage{times}
\usepackage{graphicx, subfigure}
\usepackage{epstopdf}
\usepackage{boxedminipage}
\usepackage{float}
\usepackage{url}
\usepackage{natbib, multirow, tabularx, float, amsmath, amssymb}

\def\hmpc{~h^{-1} {\rm Mpc}}

\title[The darkness that shaped the void]{The darkness that shaped the void: dark energy and cosmic voids}

\author[E.G.P. Bos et al.]{\parbox{\textwidth}{ E. G. Patrick Bos$^1$\thanks{Email: pbos@astro.rug.nl}, Rien van de Weygaert$^1$, Klaus Dolag$^{2,3}$, 
Valeria Pettorino$^4$}
\vspace*{4pt}\\
$^1$Kapteyn Astronomical Institute, University of Groningen, P.O. Box 800, 9700 AV Groningen, The Netherlands \\
$^2$Department of Physics, Ludwig-Maximilians-Universit\"at, Scheinerstr. 1, 81679 M\"unchen, Germany\\
$^3$Max Planck Institut f\"ur Astrophysik, P.O. Box 1317, D-85741 Garching, Germany\\
$^4$Universit\'e de Gen\`eve, D\'epartement de physique th\'eorique, 24 Quai Ansermet, 1211, Gen\`eve, Switzerland}

\begin{document}

\pagerange{\pageref{firstpage}--\pageref{lastpage}} \pubyear{2012}

\maketitle

\label{firstpage}

\begin{abstract}
\textit{Aims}: We assess the sensitivity of void shapes to the nature of dark
energy that was pointed out in recent studies. We investigate whether
or not void shapes are useable as an observational probe in galaxy redshift
surveys. We focus on the evolution of the mean void ellipticity and its
underlying physical cause.
\textit{Methods}: We analyse the morphological properties of voids in five sets
of cosmological N-body simulations, each with a different nature of dark energy.
Comparing voids in the dark matter distribution to those in the halo population,
we address the question of whether galaxy redshift surveys yield sufficiently accurate void morphologies. Voids are identified using the parameter free Watershed Void Finder.
The effect of redshift distortions is investigated as well.
\textit{Results}: We confirm the statistically significant sensitivity of voids
in the dark matter distribution.
We identify the level of clustering as measured by $\sigma_8(z)$ as the main
cause of differences in mean void shape $\langle\epsilon\rangle$. We find that
in the halo and/or galaxy distribution it is practically unfeasible to
distinguish at a statistically significant level between the various cosmologies
due to the sparsity and spatial bias of the sample.
\end{abstract}



\begin{figure}
 \includegraphics[width=\columnwidth]{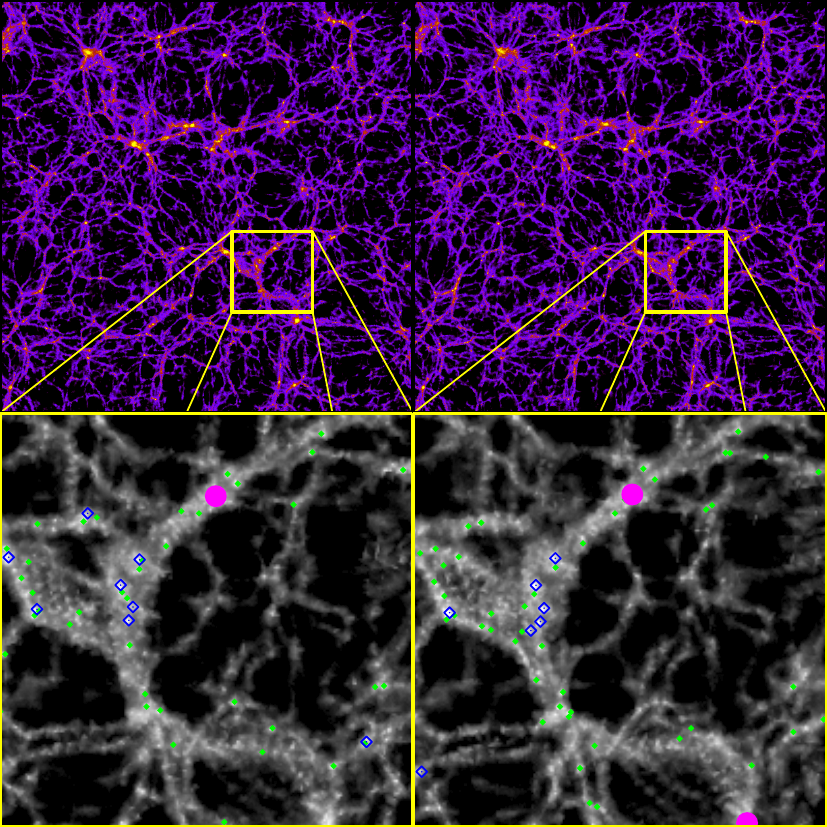}
 \caption{$\Lambda$CDM (left) and quintessence (right) cosmologies; dark matter density (grey) and galaxies (coloured dots). Differences are subtle; measuring them is the challenge of this paper.}
 \label{fig:haloes}
\end{figure}

\section{Introduction}
\label{sec:intro}

One of today's greatest puzzles is that of the nature of the dark components of our universe: dark matter (DM) and dark 
energy (DE). According to concordance cosmology, these make up respectively $21.7$\% and $73.8$\% of the total cosmic energy 
content \citep{komatsu11}, a total of $95.5$\%. Yet, there are no compelling observational indications for 
the precise nature of dark matter and dark energy \citep{komatsu11, amanullah10}.

While there is considerable agreement on some of the physical properties of dark matter, the nature of dark energy remains a
complete mystery. It is not even sure whether it really concerns an energy component to be associated to a new species in the
universe or rather a modification of gravity itself, as in extensions of general relativity like scalar-tensor theories.

It is far from trivial to constrain the nature of dark energy, due to the relatively weak imprint of the 
equation of state of dark energy, in combination with sizeable observational errors \citep{frieman08}. Nonetheless, since 
the Supernova Cosmology Project and the High-z Supernova Search Team first hinted to the accelerated expansion of the universe 
\citep{riess98,perlmutter99}, a range of astrophysical probes have led to a considerable narrowing of the parameter space 
for the nature of dark energy. Weak gravitational lensing by the large scale matter distribution, baryonic acoustic oscillations (BAOs) 
and the redshift dependent number density of clusters of galaxies are well-known examples of dark energy probes. Based on these, a 
range of observational and experimental programs have been initiated in order to constrain the nature of dark matter and dark energy. 
Notable examples are BAO experiments like WiggleZ \citep{drinkwater10}, BOSS \citep{ross10}, the 
Dark Energy Survey and large weak lensing surveys like KiDS \citep{dejong12}. Ambitious projects like ESA's 
Euclid Mission \citep{laureijs11} and the Large Synoptic Survey Telescope (LSST) will considerably sharpen and extend our knowledge on the 
dark energy content of the universe. Most of these probes directly relate to the cosmic distance measure.

Probes like the integrated Sachs Wolfe effect (ISW) \citep{dupe11} and the structure formation growth rate \citep{linder06,guzzo08}
offer alternative paths towards constraining dark energy. However, these probes are still ridden by considerable uncertainties. 
In preparation for the upcoming large dark energy experiments, it is therefore of great importance to find a wider range of 
reliable and independent probes for the determination of the dark energy equation of state. The combination of different and 
independent observations and measures of dark energy is crucial for breaking degeneracies and constraining the allowed parameter space. 
The recent study by \citet{amanullah10} illustrates the potential of such an approach to further pin down dark energy and other 
cosmological parameters.

\subsection{Voids \& cosmology}
\label{sec:voidscosmo}
Here, we are particularly interested in the imprint of dynamical dark energy \citep{wetterich88}.

Several recent studies have pointed out that cosmic voids 
not only represent a key constituent of the cosmic mass distribution, but that they are also one of the cleanest probes and 
measures of global cosmological parameters. Their structure, morphology and dynamics reflects the nature of dark energy 
\citep{parklee07, leepark09, lavaux10, biswas10, lavaux11, shoji12}, dark matter \citep{hellwing10, li11} and that of the 
possibly non-Gaussian nature of the primordial perturbation field \citep{kamionkowski09}. \citet{lavaux10} demonstrated the extreme 
sensitivity of the evolving ellipticity of voids to the equation of state of dark energy. \citet{biswas10} even quoted the 
possibility of improving the Dark Energy Task Force figure of merit by a factor of a hundred for future experiments like 
Euclid.

Along a related and perhaps even more promising route, \citet{lavaux11} demonstrated the potential of using an Alcock-Paczysnki 
test on the average shape of stacked voids \citep[also see][]{shoji12}. By stacking a large number of voids, one expects a spherical 
shape, so that size differences in radial and transverse direction can be directly related to the product of angular diameter distance 
and Hubble parameter. The claim is that for Euclid-like surveys stacked voids would outperform BAOs by an order 
of magnitude in accuracy \citep{lavaux11}.

Other studies have explored the possibility to use voids for inferring information on the amount and nature of dark 
matter. Void outflow velocities and the accompanying redshift distortions might be used to determine $\Omega_m$, the 
mass density of the universe, and infer the amount of dark matter \citep{martel90, dekel94, ryden96}. However, void outflow 
velocities are difficult to measure, while void redshift distortions are restricted because of the naturally limited void outflow. 
Nonetheless, recent advances in the study of cosmic flows, have shown the prominent and clearly recognizable dynamical role of 
voids in the nearby universe \citep{courtois12, tully08}. Related observations have been made with respect to the sensitivity 
of voids to the identification of dark matter. The emptiness of voids would be a direct measure of the strength and screening length of 
the class of long-range scalar-interacting dark matter models that have been forwarded as a possible means of explaining several 
deficiencies of the concordance $\Lambda$CDM model \citep{farrar04, gubser04, nusser05, peebles10, hellwing10, li11}. Most 
prominent amongst these deficiencies is the observed dearth of dwarf galaxies in nearby voids \citep{peebles01}. Given the extreme 
environment of voids, probing the tail of the cosmic density and halo mass distribution \citep{platen12}, they form a natural 
resort for exploring the imprint of possible modifications of general relativity, such as $f(R)$ gravity \citep{li11} and 
MOND/TeVeS \citep{llinares11}. 

\subsection{Voids}
\label{sec:voids}
Voids form a prominent aspect of the megaparsec distribution of galaxies and matter \citep{chincarini75, gregory78, einasto80, kirshner81, 
kirshner87, delapparent86, weygaert91, colless03, tegmark04, furlanetto06, huchra12}.
They are enormous regions with sizes in the range of $20-50\hmpc$ that are practically devoid of any galaxy. They are usually roundish in 
shape and occupy the major share of space in the universe \citep[see][for a recent review]{weygaert11}. Surrounded by 
elongated filaments, sheetlike walls and dense compact clusters, they weave the salient weblike pattern of galaxies and matter 
pervading the observable universe.

Voids in the galaxy distribution account for about 95\% of the total volume \citep{joeveer78, kauffmann91, elad96, elad97,
rojas05, pan12}. The typical sizes of voids in the galaxy distribution depend on the galaxy 
population used to define the voids. Voids defined by galaxies brighter than a typical $L_*$ galaxy tend to have diameters 
of order $10-20\hmpc$, but voids associated with rare luminous galaxies can be considerably larger; diameters in the range 
of $20-50\hmpc$ are not uncommon \citep{hoylevogeley02, plionis02}. Even larger voids can be recognized in the 
distribution of clusters \citep{bahcall88, einasto94, einasto01}.

Evolving out of primordial underdensities, voids become increasingly isotropic objects \citep{icke84} with a ``bucket-shape'' density 
profile whose density in the centre has a characteristic value of $\delta\approx -0.8$ \citep{sheth04}. To a first approximation, 
(isolated) spherical underdensities will become more spherical as they expand \citep{icke84,weygaert93}. In reality, voids will never 
reach perfect sphericity. Their flattening is a result of large scale dynamical and environmental influences \citep{platen08}. They 
will encounter surrounding structures such as overdense filaments or walls. Moreover, they retain an uncommonly large 
sensitivity to the dynamical influence of their large scale environment. In most situations this remains the dominant factor, 
to the extent that voids are found to become more anisotropic as time proceeds. Under realistic circumstances, the evolution 
of voids appears to reverse the simple spherical trend expected for isolated voids \citep{icke84}.

The sensitivity of voids to global cosmological parameters is a result of their unique dynamical status. On the one hand, the 
dynamical evolution of voids is rather straightforward, in that they tend to evolve into expanding, extended, uniform, and 
underdense regions with a distinct bucket-shaped profile \citep{sheth04}. On the other hand, they are distinctly 
nonlinear objects that mark the transition scale between linear and nonlinear evolution \citep{sahni94}. As such, 
their structure and morphology reflects and magnifies cosmological differences present in the primordial universe. Also, unlike 
the majority of evolving overdensities in the form of dark matter haloes and galaxies, their evolution retains the dominant 
influence of the inhomogeneous cosmological surroundings \citep{platen08}. All these factors conspire to make voids into 
important and optimal cosmological sources of information.
 
\subsection{Observing voids}
\label{sec:observingvoids}
The use of voids as a cosmological probe involves at least three major complicating factors. 

A first point of attention is the very definition of a void. This involves the need for a clear and unambiguous 
detection and delineation of the near empty void regions. A range of studies have forwarded a large and diverse number of
techniques and methods to accomplish this. For a complete overview of these algorithms we refer to the comparison study by 
\citet{colberg08} and to the study of \citet{lavaux10}. We will use the Watershed Void Finder, developed by our group 
\citep{platen07}. This method allows a parameter-free determination of the size and shape of voids in the matter and 
galaxy distribution. We will discuss some of the relevant technical details in \S\ref{sec:voidID}. 

A second issue concerns the fact that the relevant observations of cosmic structure consist almost exclusively of 
galaxy redshift surveys. The use of redshifts for distance estimation distorts the observed shapes of structures \citep{shoji12}. 
This is a consequence of the peculiar velocities in and around these objects. In the case of voids, matter and galaxies are flowing 
out of the density depressions as a result of the lower than average gravitational attraction. As a result, with respect to the 
Hubble flow we see galaxies at the frontside moving towards us and those at the backside moving away from us, stretching the void 
along the radial direction. A number of studies in the nineties proposed this as a possible means of extracting global cosmological 
information from voids \citep[e.g][]{ryden96}. Recent studies confirm this idea \citep{percival09, jennings11}.

A third and major complicating factor for the study of voids in observations is the fact that we have to infer 
their properties from the diluted and possibly strongly biased population of galaxies.
Galaxy redshift surveys are necessarily limited in their spatial resolution of cosmic 
structure. Even in the Sloan Digital Sky Survey (SDSS), structure is delineated by relatively bright galaxies. The 
spatial void distribution defines an intrinsically hierarchical complex of voids, in which a void consists 
of an assembly of subvoids \citep{sahni94, sheth04, aragon10, aragon12}. Finer substructure can only be observed 
when sampling ever fainter objects. Hence, the brighter galaxies of SDSS only trace the rough outline of 
larger voids, with sizes in the order of $10-15\hmpc$. Even more complicating is the issue of the possible 
bias of galaxies with respect to the dark matter distribution. While moderate density regions ($\delta\sim 0-$ a few) 
appear to have 
a rather low level of bias, a range of studies have indicated a strong 
bias of galaxies in and around the lowest density regions. \citet{peebles02} strongly emphasized the fact that 
dwarf galaxies seem to avoid the void regions, contrary to the expectations and predictions of most galaxy 
formation theories. This \textit{void phenomenon} is one manifestation of what appears to be a 
rather puzzling situation with respect to the galaxy distribution in voids: different semi-analytical schemes of 
galaxy formation predict either bias or anti-bias \citep{platenphd,platen12}. 

In figure~\ref{fig:haloes} the galaxies/haloes (points) are overlaid on the full density field, illustrating that 
the spatial distribution of the two don't match. This is an especially poignant example of the fact that void substructure 
is lost on small scales. The large void of $\sim 30 \hmpc$ at the right of the figure is traced only on the outside by 
galaxy-like ``haloes'' (more on this in \S\ref{sec:dm_haloes}).

\subsection{Voids in the halo distribution}
\label{sec:voidhalo}
In this study, we address the issue of whether it is feasible to infer the nature of dynamical dark energy 
from the evolving population of voids in the observational context of the voids being sampled by a dilute 
population of discrete objects, like dark matter haloes and galaxies. 

In this, we follow the strategy of identifying voids only and solely on the basis of the spatial 
distribution of the sampled objects. From the distribution of galaxies or haloes, we outline voids 
with the help of the Watershed Void Finder (WVF). While under ideal circumstances, one might consider 
to use knowledge of the dynamical evolution of voids, we chose not to do so given the uncertainties 
about the bias of void galaxies: it would amplify any difference between the sampled halo/galaxy 
distribution and the underlying density field. Our plan is to follow the evolution of the shape of 
voids inferred from the spatial \textit{dark matter} and \textit{dark halo} distribution at each epoch and to compare 
the results obtained for different dynamical dark energy models. The void population in the \textit{dark matter} distribution 
of the different dark energy cosmologies is expected to reflect the sensitivity to the nature of dark energy. How 
accurately the void population in the \textit{dark halo distribution} manages to follow this sensitivity, is one 
of the main questions to be answered by this study.

Also, we investigate the origin of the dependence of voids to the nature of dark energy. We 
presume that the structure of voids at any cosmic epoch is a reflection of the stage of structure 
development in the corresponding dark energy cosmology. This suggests a strong relation between 
the amplitude of the density and velocity perturbations in the mass distribution and the structure, 
shape and profile of the voids. We therefore assess in how far the void shape evolution in different 
dark energy scenarios can be ascribed to $\sigma_8$, the root mean square density fluctuation within 
a sphere of radius $8\hmpc$, and whether there are other possible factors of influence. 

\subsection{Outline}
\label{sec:outline}
The outline of this paper is as follows. In \S\ref{sec:theory} we will treat cosmological models, four of which contain a time dependent 
dark energy component and discuss cosmic voids and their theoretical use in discerning between cosmological models. We will give a 
description of the data in \S\ref{sec:data}. Void identification is invaluable to our analysis and we will elaborate on this in 
\S\ref{sec:voidID}. In \S\ref{sec:analysis_void} we describe the void populations in the simulations of the various dark energy 
scenarios, along with their size and shape characteristics. Subsequently, in \S\ref{sec:shapeVSz} we examine the evolution of 
the void shapes, and try to assess under which conditions these can be used as probes, given the discrete, diluted and biased 
halo and galaxy samples on which the analysis is based. We finish our analysis in \S\ref{sec:sigma8}, where we touch on the 
issue of clustering dependence. We conclude and discuss our findings in \S\ref{sec:discussion}. 

\section{Dark energy \& voids}
\label{sec:theory}
\subsection{Dark energy cosmology}
\label{sec:cosmology}
The precise nature of dark energy will be a decisive factor for the fate of the universe.
Our reference point in comparing different cosmological models is that of ``concordance'' cosmology or $\Lambda$CDM 
cosmology. The geometry of the universe is flat and its matter content is dominated by a species of 
cold dark matter, while baryonic matter represents a smaller proportion. Its dynamics 
is currently dominated by the cosmological constant $\Lambda$.

In this paper, we are going to investigate a range of possible dynamical dark energy alternatives to the cosmological 
constant. The nature of dark energy is often specified in terms of the equation of state, $w=P/\rho$. The first order 
parametrization of its equation of state, 
\begin{equation}
\label{eqn:DEgeneralparam}
w(a) = w_0 + w_a(1-a) \,,
\end{equation}
is often used to compare different dark energy models \citep{chevallier01, linder03}.
The impact of the nature of the -- possibly dynamic and evolving -- dark energy on the evolution of the 
universe, follows directly from the Friedmann equation, 
\begin{equation}
	\label{eqn:friedmann}
	\left(\frac{H}{H_0} \right)^2 = \sum_j \Omega_{0,j} \exp\left( -3 \int_{a_0}^a \frac{1+w_j(a')}{a'} da' \right) \,.
\end{equation}
In this expression the Hubble parameter is $H \equiv \dot{a}/a$, $a(t)$ is the (normalised) cosmological scale factor and the dot means derivative with respect to cosmic time. 
The relative energy densities of the cosmological fluids -- such as baryons, dark matter, radiation and dark energy -- at 
the current cosmological epoch is given by $\Omega_{0,j}$. For a flat universe, we evidently have $\sum_j\,\Omega_j(t)=\sum_j\,\Omega_{0,j}=1$. 

\begin{figure}
\includegraphics[width=\columnwidth]{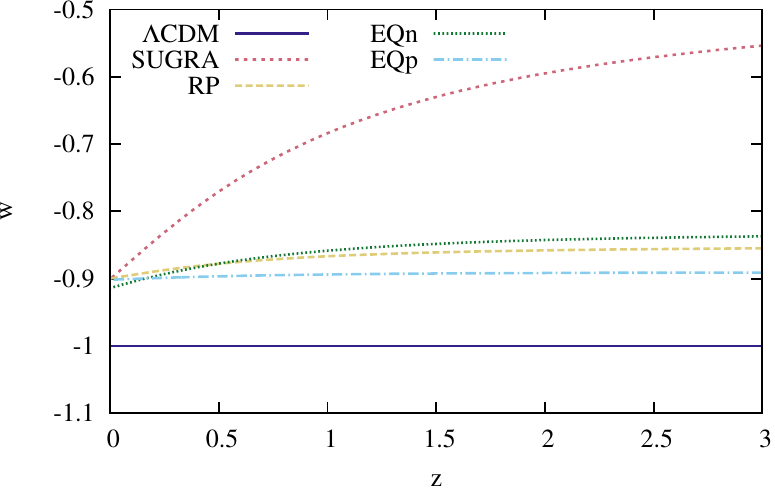}
\caption{Equation of state parameter $w$ versus redshift. Shows the evolution of the dark energy equation of state parameter $w = P/\rho$. The different lines indicate different dark energy models.}
\label{fig:wvsz}
\end{figure}

\subsection{Dynamical dark energy}
\label{sec:dynDE}
There are two enigmas if dark energy is in fact to be ascribed to a cosmological constant $\Lambda$. The first riddle concerns 
the realisation that it is relatively unlikely for the measured densities of matter and dark energy to be almost equal at the 
present time, because -- according to a theory with a cosmological constant -- they must have been very different in the past 
and they will be very different in the future. This is aptly called the 
\textit{coincidence problem}, and concerns the question why the density of matter and dark energy are of comparable magnitude at the present 
epoch. The second puzzle concerns the embarrassingly small value of the cosmological constant. 
Particle physics tells us that there should be a vacuum energy, and that this would have the effect of a cosmological constant. However, the 
predicted value of the energy density of the vacuum is about 120 orders of magnitude higher than what is measured for $\Omega_\Lambda$.
If the interpretation of the cosmological constant as a form of vacuum energy would be right, this would call for a fine 
tuning of the initial conditions over an unconceivable large dynamic range. This is known as the \textit{fine-tuning problem}.

Dynamical dark energy has been proposed in order to find a possible and more natural solution \citep{wetterich88,rp88}.
The most interesting cases are obtained in the presence of attractor solutions \citep{liddle99}, in which the
dark energy scalar field follows a trajectory which is the same for a wide range of initial conditions of the scalar field and its
derivative. Depending on the potential, dark energy can track the background component and then dominate over it at relative 
recent times \citep{copeland98}.

In this work we investigate four cosmologies with different time dependent dark energy models. The first two models are basic quintessence models: 
the Ratra-Peebles (RP) and SUGRA models. These models involve a scalar field that does not interact with the other cosmic species, 
except minimally through the overall cosmic expansion. The difference between RP and SUGRA is in the potential $V(\phi)$ that 
describes the dynamics of the scalar field $\phi$. The other two models are extended quintessence (EQ) models. Their Lagrangian 
contains a term which is responsible for a non-minimal coupling of the scalar field to gravity. Effectively, this introduces a 
``fifth force'' between matter particles which modifies their interaction \citep{baccigalupi00, perrotta02, pettorino05a, pettorino05b, 
pettorino08}. Effectively, the extended quintessence models are scalar-tensor theories of gravity \citep{hwang90a, hwang90b, faraoni00, 
boisseau00, esposito01, riazuelo02}. The coupling constant can be either positive or negative, with different effects. We model both 
possibilities; the model with negative constant is called EQn and the one with positive constant is called EQp. In addition to these 
four cosmologies, we also investigate the standard $\Lambda$CDM cosmology. 

The differences between the five dark energy models can be appreciated on the basis of the evolution of their equation of state parameter, $w(z)$, 
which is shown in figure~\ref{fig:wvsz}. The largest differences, over the full redshift span, are those between $\Lambda$CDM and 
SUGRA. At $z=2$, the difference is no less than 0.4. At $z=0.5$ it has decreased to 0.2. The technical details on the models are presented 
in appendix~\ref{sec:modelsofDE} and in \citet{deboni11}.

Minimally coupled quintessence models, even including a tracking behaviour, still suffer from the problem of fine tuning of the initial conditions, 
because the observational bounds on the dark energy equation of state are increasingly converging towards a value of $w_\mathrm{DE}$ very close to $-1$. 
The closer $w_\mathrm{DE}$ is to the cosmological constant value, the smaller the range of allowed initial values for the corresponding scalar field 
has to be, since the dynamics of such a field is extremely constrained by the flatness of the potential in which the field evolves \citep{matarrese04, pettorino05b}. This is also the reason why people have been investigating further alternatives, often requiring extensions of general relativity
which modify the gravitational attraction felt by matter particles. Two popular ways of proceeding
are either modifying the coupling with gravity itself, in the Jordan frame, as in scalar-tensor,
EQ, theories or to change the coupling of dark matter particles only, directly in the Einstein frame 
\citep[and references therein]{amendola00, pettorino08}.

In this paper we follow the first path. In EQ cosmologies, the non-minimal coupling to gravity
induces an ``R-boost'' mechanism responsible for early, enhanced scalar field dynamics, by virtue
of which the residual imprint of a wide set of initial field values is cancelled out. These
models, therefore, `extend' the attractor solution behaviour of quintessence fields to scalar-tensor
cosmologies. However, we note here that even in these cases a flat potential is still
required in order to get a reasonable equation of state today. Also, scalar-tensor cosmologies (EQ
and $f(R)$) involve a coupling to baryons too, which therefore require the presence of some
`chameleon'-like mechanism that protects the mass of the scalar field in high density regions.
Nevertheless, this set of models has interesting solutions (attractors) and is simple to
implement, therefore being a good candidate to test differences with quintessence models through
N-body simulations.

\subsection{Void shapes \& dark energy}
\label{sec:voidDE}
Following the earlier suggestions by \citet{parklee07,leepark09}, recent studies by Wandelt and collaborators 
\citep{lavaux10,biswas10} showed that voids may be used as precision probes of dark energy. 

The sensitivity of voids to dark energy is a result of the way in which the dark energy equation of 
state affects the dynamical evolution of voids via its imprint on the large scale tidal force fields. 
Their influence remains important during the entire evolution of voids. The tidal forces evoke a 
significant anisotropic effect in the development of the voids, even sometimes causing their complete 
collapse \citep{sheth04}. 

As a result, the elliptical shape parameters that describe the flattening or elongation of a void are expected 
to be intimately connected to the local tidal tensor
\citep[see equations~\ref{eqn:tideig1} \& \ref{eqn:tideig2} and][]{bond96, parklee07, platen08}.
It relates their 
shapes directly to the surrounding inhomogeneous cosmic matter distribution responsible for the gravitational tidal 
field. In turn, the evolution of the tidal forces are directly dependent on the nature of the dark energy content 
of the universe.
This offers a path towards measuring the cosmological parameters. 

The dependence of the void's sphericity $s$ and oblateness $p$ (equation~\ref{eqn:shape_parameters}) on the tidal tensor $T_{ij}$, defined as the traceless 
component of the second derivative of the gravitational potential $\phi$,
\begin{equation}
T_{ij} = \frac{\partial^2\phi}{\partial x_i\partial x_j} - \frac{1}{3}\nabla^2\phi\,\delta_{ij} \,.
\label{eqn:tide}
\end{equation}
can be directly inferred from their relation to the (ordered) eigenvalues of the tidal tensor,  
$\lambda_1 > \lambda_2 > \lambda_3$:
\begin{equation}
\lambda_1(p,s) = \frac{1+(\delta_v-2)s^2 + p^2}{p^2 + s^2 +1}
\label{eqn:tideig1}
\end{equation}
\begin{equation}
\lambda_2(p,s) = \frac{1+(\delta_v-2)p^2 + s^2}{p^2 + s^2 +1} \,,
\label{eqn:tideig2}
\end{equation}
where $\delta_v = \sum_{i=1}^3\lambda_i$. We can infer the dependence of the ellipticity distribution of voids on the cosmological model. 
The probability density distribution $f(s)$ for the sphericity $s = 1 - \epsilon$, with $\epsilon$ the ellipticity, is as follows:
\begin{equation}
\label{eqn:parkleeProb}
\begin{split}
&f(1-\epsilon; z) = f(s; z, R_L) = \int_s^1 \mathcal{P}[p,s|\delta=\delta_v; \sigma(z, R_L)]dp \\
&= \int_s^1 dp \frac{3375\sqrt{2}}{\sqrt{10\pi}\sigma^5(z,R_L)}\exp\left[ \frac{-5\delta_v^2 + 15\delta_v(\lambda_1+\lambda_2)}{2\sigma^2(z,R_L)} \right] \\
& \times \exp\left[ -\frac{15(\lambda_1^2+\lambda_1\lambda_2 + \lambda_2^2)}{2\sigma^2(z,R_L)} \right](2\lambda_1+\lambda_2 - \delta_v)\\
& \times (\lambda_1 - \lambda_2) (\lambda_1 + 2\lambda_2 - \delta_v) \frac{4(\delta_v - 3)^2ps}{(p^2 + v^2 +1)^3}  \,.
\end{split}
\end{equation}

This distribution is sensitive to changes in the cosmological parameters \citep{parklee07} through $\sigma(z, R_L)$, the linear rms 
fluctuation of the matter density field smoothed on a scale of $R_L$ at redshift $z$, defined as:
\begin{equation}
\label{eqn:sigma_general}
\sigma^2(z, R_L) \equiv D^2(z)\int_0^\infty \frac{k^2 dk}{2\pi^2} P(k) W^2(kR_L) d\ln k \,,
\end{equation}
where $D(z)$ is the linear growth factor, $W(kR_L)$ is the top-hat window function and $P(k)$ the linear power 
spectrum. The filter scale $R_L$ is directly related to the void size in Lagrangian space, 

Equation~\ref{eqn:parkleeProb} implies that the mean ellipticity of voids decreases with redshift $z$, where the mean ellipticity 
$\langle\epsilon\rangle$ is defined as
\begin{equation}
	\langle\epsilon\rangle = \int \epsilon\, f(\epsilon,z) d\epsilon \,.
\end{equation}
More importantly, the model's rate of ellipticity decrease is sensitive to changes in the cosmological parameters. The redshift dependence 
of the mean ellipticity can, thus, be used to discriminate between different values of $w_a$ \citep{leepark09, lavaux10}. 

These results support the impression that voids are a promising probe of the nature of dark energy. Indeed, as \citet{biswas10} showed, 
we can improve upon our constraints on cosmological parameters by including void shape data, e.g.\ from Euclid.
One of the central tests of this study is to see whether indeed we can use this probe in our simulations.

\section{Simulations \& Samples}
\label{sec:data}
Our analysis is based on numerical N-body simulations of DM particles in five different cosmological backgrounds, four of which include evolving 
dark energy. We use version 3 of the \texttt{GADGET} code \citep{springel05}, which has the ability to specify the mode of evolution of 
dark energy. This is achieved through an extended dark energy implementation, described in \citet{dolag04}.

We simulate boxes with periodic boundary conditions. The physical linear size of a box is $300 \hmpc$. They contain DM particles only. 
The initial conditions at $z=60$ are set up using the Zel'dovich approximation \citep{zeldovich70}.
The general cosmological parameters adopted for the simulations are the WMAP 3-year data values: $\Omega_m = 0.268$, 
$\Omega_\Lambda = 0.732$, $\Omega_b = 0.044$, $h = 0.704$, $\sigma_8 = 0.776$ and $n = 0.947$. We use two different mass resolutions ($256^3$ 
and $768^3$ particles) for our simulations and build DM halo catalogs from the $768^3$ sets.

We simulate five different dark energy models (\S\ref{sec:dynDE}). The specific model parameters are summarised in table~\ref{tab:DEparams}. 
These parameters are consistent with current observational constraints \citep{acquaviva05, amanullah10, komatsu11}. The tables of $w(a)$, 
needed for the extended dark energy implementation, are calculated for the different models (see figure~\ref{fig:wvsz}).

\begin{table}
\begin{tabular*}{\columnwidth}{@{\extracolsep{\fill}}lrrrrrr}
\itshape Model & \itshape $\alpha$& \itshape $\xi$& \itshape $w_{JBD}$& \itshape $w_0$& \itshape $w_a$& \itshape $\sigma_8$ \\ \hline
$\Lambda$CDM & -- & -- & -- & $-1.0$ & $0.0$ & $0.776$ \\
RP & $0.347$ & -- & -- & $-0.9$ & $0.0564$ & $0.746$ \\
SUGRA & $2.259$ & -- & -- & $-0.9$ & $0.452$ & $0.686$ \\
EQp & $0.229$ & 0.085 & $120$ & $-0.9$ & $0.0117$ & $0.748$ \\
EQn & $0.435$ & -0.072 & $120$ & $-0.9$ & $0.0805$ & $0.729$ \\  \hline
\end{tabular*}
\caption{Dark energy model parameters used in the simulations. For a description of the models see \S\ref{sec:modelsofDE}.
$\alpha$ is the power-law slope of the quintessence potential, $\xi$ is the coupling constant for the extended quintessence
models and $w_{JBD}$ is given by equation~\ref{eqn:wJBD}. $w_0$ is fixed at $z=0$.
$w_a$ is determined using a $\chi^2$ fit (see equation~\ref{eqn:DEgeneralparam}) and $\sigma_8$ is given at $z=0$ ($\sigma_8$
is the same for all models at $z\simeq1089$).}
\label{tab:DEparams}
\end{table}

\begin{figure}
\includegraphics[width=\columnwidth]{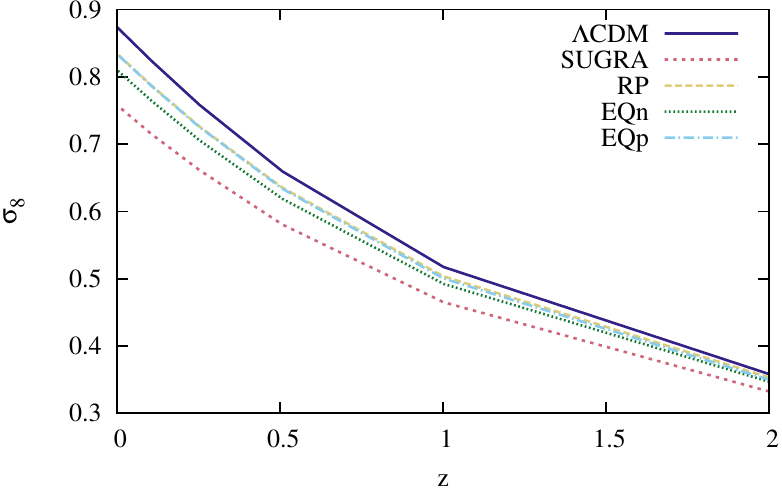}
\caption{$\sigma_8$ as a function of redshift. The standard deviation of the values of the density field on a scale of $8\hmpc$, $\sigma_8$, is a measure of the amount of clustering in the field. We show the values of $\sigma_8$ for the $\Lambda$CDM and four quintessence models as a function of redshift. This is an indicator of the evolution of structure in the different cosmologies.}
\label{fig:sigma8vsz}
\end{figure}

The dark energy models are normalised at the CMB using the relation
\begin{equation}
\label{eqn:sigma8norm}
\sigma_{8,\mathrm{DE}} = \sigma_8 \frac{D_\mathrm{{\Lambda}CDM}(z_\mathrm{CMB})}{D_\mathrm{DE}(z_\mathrm{CMB})} \,,
\end{equation}
where we assume $z_\mathrm{CMB} = 1089$ and $D$ is the linear growth factor, which is dependent on the dark energy model through $H$. This rescaling 
will cause differences in the amount of clustering, characterised by the normalisation parameter $\sigma_8$. In figure~\ref{fig:sigma8vsz} we 
show the resulting measured values of $\sigma_8$ for five different models (taken from our own low resolution data, see below). It is clear 
that the dark energy models have their influence on structure formation. In \S\ref{sec:analysis_void} and \S\ref{sec:shapeVSz} 
we will see that its impact is significant. We describe these samples in more detail below.

\subsection{DM particles}
\label{sec:dm_particles}
For each one of the five different dark energy models (\S\ref{sec:dynDE}), we use a high resolution cosmological N-body simulation 
of $768^3$ DM particles. Each of these simulations has exactly the same initial conditions. The particles have masses of $0.443 \times 10^{10} h^{-1} M_\odot$. 
For the $\Lambda$CDM model we have snapshots at $z = 0.1$, $z=0.25$, $z=0.51$, $z=1.00$ and $z=2.04$, while for the other models we have a 
high resolution snapshot at $z=0$. More information on these data can be found in \citet{deboni11}.

For each of the five DE models we also generated and computed eight low-resolution simulations, each with $256^3$ DM particles. 
The physical parameters used for the initial conditions of these sets are exactly the same as those of the high resolution 
simulations. The lower resolution simulations are used to investigate the time evolution of the dark energy models and to investigate the 
influence of mass resolution on the statistics used in our analyses (\S\ref{sec:shapeVSz}). By using different random realisation for 
the initial conditions, these simulations are also essential for assessing and ruling out possible cosmic variance effects (\S\ref{sec:error}).

In addition, we obtain random particle samples from the high resolution simulation sets, each sample containing the same 
number of particles as the number of dark matter haloes. These samples are used to evaluate the effect of biasing of 
the halo population (\S\ref{sec:shapeVSz}).

\subsection{DM haloes}
\label{sec:dm_haloes}
For every high resolution simulation box we use the \texttt{SUBFIND} algorithm of \citet{springel01} to find the gravitationally bound 
haloes (groups of adjacent particles, representing concentrated clumps of DM) that can colloquially be identified as galaxy haloes.

\begin{figure*}
\includegraphics[width=\textwidth]{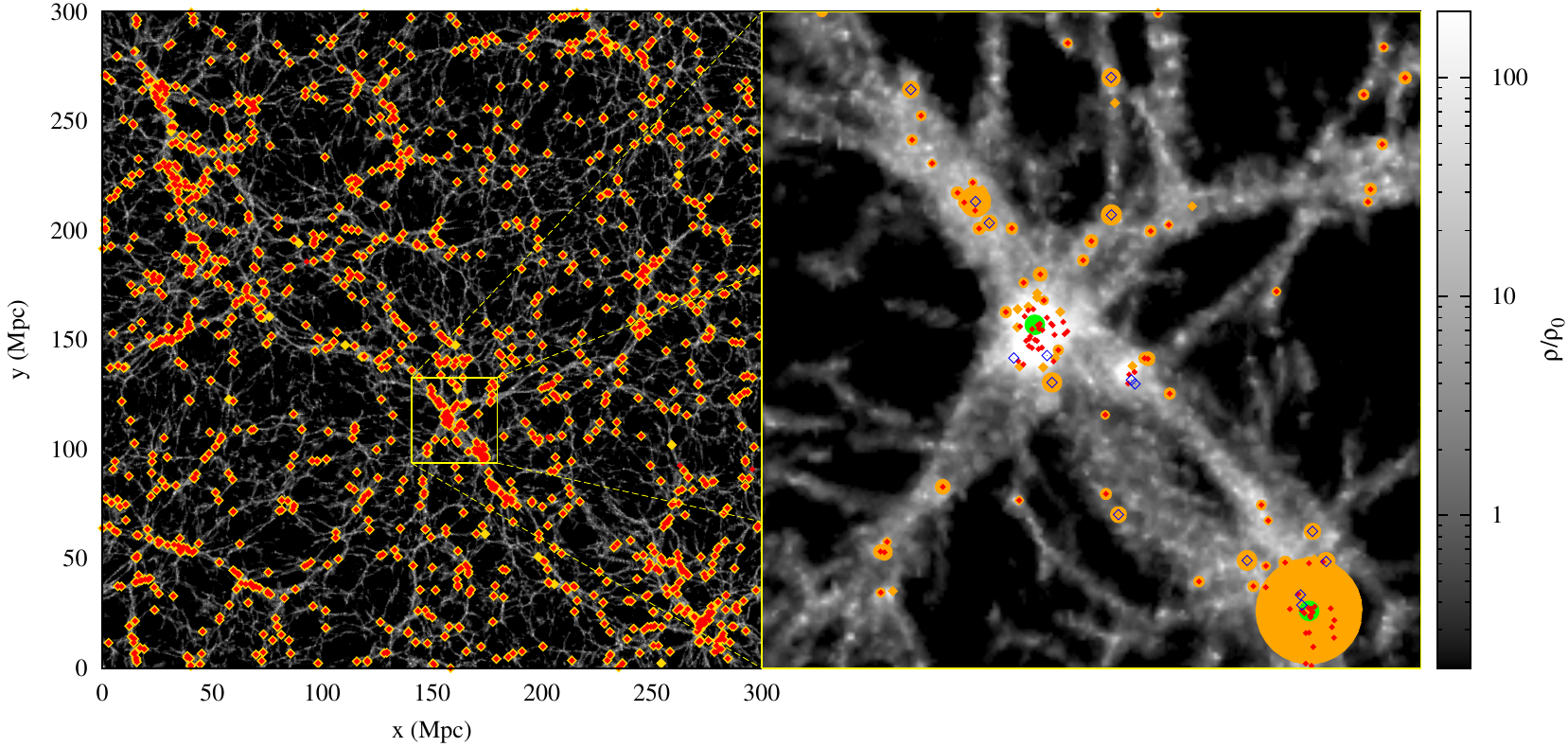}
\caption{Halo distribution.
In the left panel we show the distribution of haloes (gold) and groups of haloes (red) in a slice of thickness $0.25\hmpc$
of the $\Lambda$CDM simulation at $z=0$. The density is shown in grayscale; the corresponding values are indicated by the color bar to the right.
In the right zoom-in panel we additionally split the haloes by mass; red dots have $M < 10^{12}M_\odot$, blue open diamonds have 
$10^{12}M_\odot < M < 10^{13}M_\odot$ and the two green filled circles are even more massive; the centre one has a mass of $2.4\cdot10^{13}M_\odot$, 
the lower right one has $3.1\cdot10^{14}M_\odot$. The orange circles indicate the location of haloes and groups. The radii of the circles 
are proportional to their mass.}
\label{fig:haloslice}
\end{figure*}

The \texttt{SUBFIND} haloes (also called subhaloes, a term we will not use) trace the general structures present in the field. 
Figures~\ref{fig:haloes} and \ref{fig:haloslice} show that haloes are found in clusters, filaments and also 
the more tenuous walls. The voids remain largely empty. Moreover, careful observation shows that the more massive haloes are found in the 
higher density cluster nodes and most pronounced filaments. This reflects the fact that the mass function of haloes is 
dependent on the large scale environment: the mass function of void haloes is shifted to low mass objects \citep[see e.g.][]{aragon07}. 

Most important for our study is the observation that while the haloes do trace the outline of the most 
prominent large scale features, the detailed substructure is lost in the halo distribution. This is immediately 
obvious when comparing the spatial halo distribution in figure~\ref{fig:haloslice} with the underling 
dark matter distribution. We may therefore expect to find a substantially different void size -- and probably also 
void shape -- distribution than in the intricate large scale structure visible in the dark matter distribution. 

\begin{table}
\begin{small}
\begin{tabular*}{\columnwidth}{@{\extracolsep{\fill}}rlrrrr}
\itshape $z$ & \itshape Model & \itshape $N_\mathrm{haloes}$& \itshape $\langle m \rangle$& \itshape $\sigma_m$& \itshape $m_\mathrm{max}$ \\ \hline
$0$  & WMAP3         & 567119        & $1.1\cdot10^2$     & $7.0\cdot10^2$     & $1.7\cdot10^5$ \\
  & SUGRA         & 582882        & $9.5\cdot10^1$     & $4.9\cdot10^2$     & $8.9\cdot10^4$ \\
  & RP    & 575618        & $1.0\cdot10^2$     & $6.0\cdot10^2$     & $9.6\cdot10^4$ \\
  & EQn   & 577526        & $1.0\cdot10^2$     & $5.6\cdot10^2$     & $9.4\cdot10^4$ \\
  & EQp   & 572176        & $1.0\cdot10^2$     & $6.0\cdot10^2$     & $9.7\cdot10^4$ \\ \hline
$1$  & WMAP3         & 469725        & $6.7\cdot10^1$     & $2.1\cdot10^2$     & $1.9\cdot10^4$ \\
  & SUGRA         & 446675        & $6.0\cdot10^1$     & $1.7\cdot10^2$     & $1.6\cdot10^4$ \\
  & RP    & 464471        & $6.5\cdot10^1$     & $2.0\cdot10^2$     & $1.8\cdot10^4$ \\
  & EQn   & 455982        & $6.3\cdot10^1$     & $1.9\cdot10^2$     & $1.7\cdot10^4$ \\
  & EQp   & 463068        & $6.5\cdot10^1$     & $2.0\cdot10^2$     & $1.8\cdot10^4$ \\ \hline
\end{tabular*}
\end{small}
\caption{Characteristics of the halo catalogues obtained from the DE simulations, at $z=0$ and $z=1$. From left to right, we show the number of haloes $N_\mathrm{haloes}$, the mean halo mass $\langle m \rangle$, the standard deviation of the halo masses $\sigma_m$ and the largest mass in that snapshot $m_\mathrm{max}$. Masses are in units of $10^{10} M_\odot$.}
\label{tab:haloes}
\end{table}

\subsubsection{Halo data}
We produce halo catalogues of all the dark energy simulations at redshifts of $z=0$, $z = 0.1$, $z=0.25$, $z=0.51$, $z=1.00$, $z=2.04$, 
$z=2.98$ and $z=3.80$. Because the halo finder finds significantly less haloes at higher redshifts, the corresponding halo samples 
are unsuitable for comparison with the lower redshift halo samples. Some basic data on the halo samples in the five different 
cosmological simulations are listed in table~\ref{tab:haloes}. 

The position of a halo is simply defined as its centre of gravity, which is the mean position of its $N$ constituent 
particles, given each has the same mass:
\begin{equation}
{\vec x} = \frac{1}{N} \sum_i^{N} {\vec x}_i \,.
\end{equation}
where ${\vec x}_i$ are the particle positions and $N$ is the number of particles in the halo. The velocity is calculated 
analogously. Halo mass is simply the sum of the particle masses.

\begin{figure}
\includegraphics[width=\columnwidth]{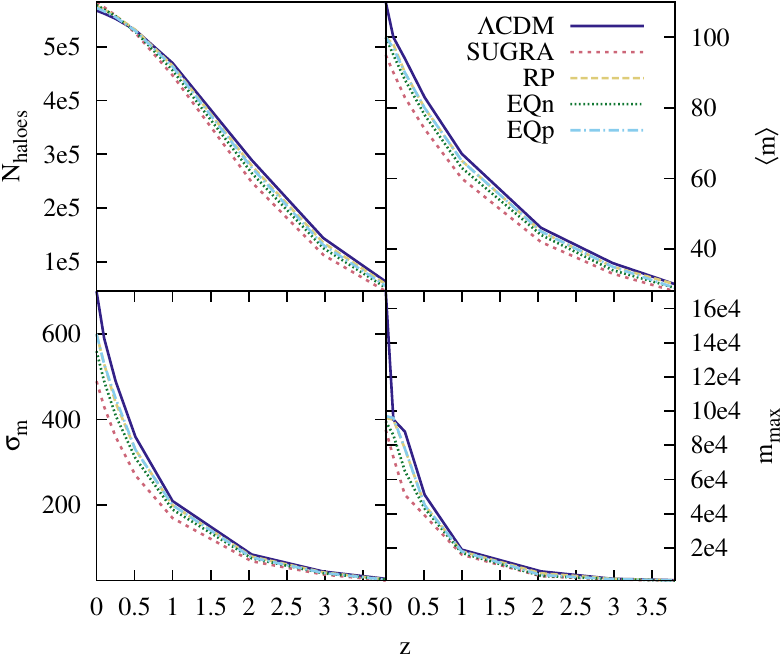}
\caption{Characteristics of the halo catalogues as a function of redshift. Different lines show catalogues obtained from the different dark energy simulations. Top left: number of haloes $N_\mathrm{haloes}$. Top right: the mean halo mass $\langle m \rangle$. Bottom left: the standard deviation of the halo masses $\sigma_m$. Bottom right: the largest mass in that snapshot $m_\mathrm{max}$. Masses are in units of $10^{10} M_\odot$.}
\label{fig:halo_data}
\end{figure}

In figure~\ref{fig:halo_data} we show some further characteristics of the halo catalogues as a function of redshift. 
The upper left panel shows the total number of haloes found at the given redshift. We observe that up until about $z \sim 0.5$ the 
number of haloes is largest in the $\Lambda$CDM cosmology. After this, the quintessence models take over. This hints at a lower rate 
of halo mergers, the main driving force behind hierarchical structure evolution. Indeed, it is mainly due to the lowest mass haloes, 
meaning that these have not been able to merge into larger haloes. Note that \citet{klypin03} conclude the opposite. This is because 
they normalise the power spectra of the different cosmologies at $z=0$ instead of at $z=1089$ as we do. In \S\ref{sec:sigma8} we 
discuss this matter further.

The other panels show (clockwise, starting at the top right) the mean halo mass $\langle m\rangle$, the maximum halo mass in the 
sample $m_\mathrm{max}$ and the standard deviation in the mass distribution $\sigma_m$. These panels show that at every redshift 
the haloes in the SUGRA cosmology are less massive than those of the others, implying that evolution of structure in the SUGRA 
universe is slowest. See \citet{deboni11} for a thorough analysis of the halo properties and their possible direct use as probes 
of the nature of dark energy.

\section{Void identification \& analysis}
\label{sec:voidID}
Our void analysis of the particle and halo samples involves a few steps. In this section 
we will describe these in some detail, along with a schedule of the analysis setup. 

The fundamental step of the entire analysis is that of finding and defining voids. For this we use the 
Watershed Void Finder \citep{platen07}. Having identified the voids, we determine the shapes and sizes of 
the voids. Subsequently, we need to identify and evaluate the possible systematic effects that 
play a role in the inferred void properties. Also, for the interpretation of the significance of 
the results, we have to take into account the cosmic variance. 

\subsection{Watershed Void Finder}
\label{sec:structure_voids}
Depending on your definition, voids can make up from 13 to 100\% of the total volume. 
Evidently, there is no unanimous agreement on the definition of \emph{void}. It is one of the reasons 
why there are a wide range of different void finding algorithms, often yielding vastly different 
results \citep[see e.g.][]{colberg08}. For our statistical analysis we use the void finder developed 
by our group, the Watershed Void Finder (WVF), introduced by \cite{platen07}. It is a largely 
parameter free formalism that manages to outline a void region, independently of its size and 
shape, and thus ideally suited for an objective statistical study. Moreover, the study by 
\cite{platen08} showed that the voids identified by WVF are indeed closely related to the physical 
structure of the mass distribution: their orientation is intimately coupled to the 
tidal field, closely following the related theoretical predictions.

The Watershed Void Finder (WVF) is an implementation of the \emph{Watershed 
Transform} for segmentation of images of the galaxy and matter distribution 
into distinct regions and objects and the subsequent identification of voids. 
The basic idea behind the watershed transform finds its origin in geophysics. It 
delineates the boundaries of the separate domains, the \emph{basins}, into which 
yields of e.g.\ rainfall will collect. The analogy with the cosmological context is 
straightforward: \emph{voids} are to be identified with the {\it basins}, while 
the {\it filaments} and {\it walls} of the cosmic web are the ridges separating 
the voids from each other. 

With respect to the other void finders the watershed algorithm has several advantages. 
Because it identifies a void segment on the basis of the crests in a density field 
surrounding a density minimum it is able to trace the void boundary even though it 
has a distorted and twisted shape. Also, because the contours around well chosen minima 
are by definition closed, the transform is not sensitive to local protrusions between 
two adjacent voids. The main advantage of the WVF is that for an ideally smoothed density 
field, it is able to find voids in an entirely parameter free fashion.

The WVF consists of eight steps, which are extensively outlined in \cite{platen07}.  
For the success of the WVF it is of utmost importance that the density field retains 
its morphological character. To this end, the two essential first steps relate 
directly to the Delaunay Tessellation Field Estimator (DTFE), which guarantees the correct 
representation of the hierarchical nature, the weblike morphology dominated by filaments and 
walls, and the presence of voids \citep{schaap00,weygaert11,cautun11a}. The obtained densities 
are interpolated onto a regular grid of $384^3$ grid cells. 

Because in and around low-density void regions the raw density field is characterised by a 
considerable level of noise, a second essential step suppresses the noise by means of
a properly defined filtering operation. In its original definition, we invoked an adaptive 
smoothing algorithm, {\it median filtering}, which in a consecutive sequence of repetitive steps 
determines the median of densities within the {\it contiguous Voronoi cell} surrounding a point. 
The determination of the median density of the natural neighbours turns out to define a stable and 
asymptotically converging smooth density field fit for a proper watershed segmentation. An 
alternative filtering operation is one in which we smooth the field by Gaussian filters of 
a fixed scale. The WVF segmentation of the filtered field produces the void population 
corresponding to that one particular scale. In principle, one can combine this with the 
WVF void segmentations at other scales. By carefully examining and linking small scale 
void structure and the large scale voids one may infer the multi-scale structure 
of the hierarchically evolved void population \citep{sheth04,aragon12}. In our study, 
we have opted for this last filtering operation, as it will allow us to explicitly assess 
the scale dependence of the void shape measurements. This will enable us to find explicit scale 
criteria for a successful dark energy exploration based on the void population. 
 
The subsequent central step of the WVF formalism consists of the application of the discrete 
watershed transform on this adaptively filtered density field. As a result of noise in the 
(discretely) sampled density field, the WVF segmentation includes artefacts. To a first order 
these can be classified as {\it false splits} and {\it false mergers} of void-patches. To deal 
with this these effects, we include a step of removal of segmentation boundaries whose density 
is lower than some user-defined density threshold.

\begin{figure*}
\includegraphics[width=\textwidth]{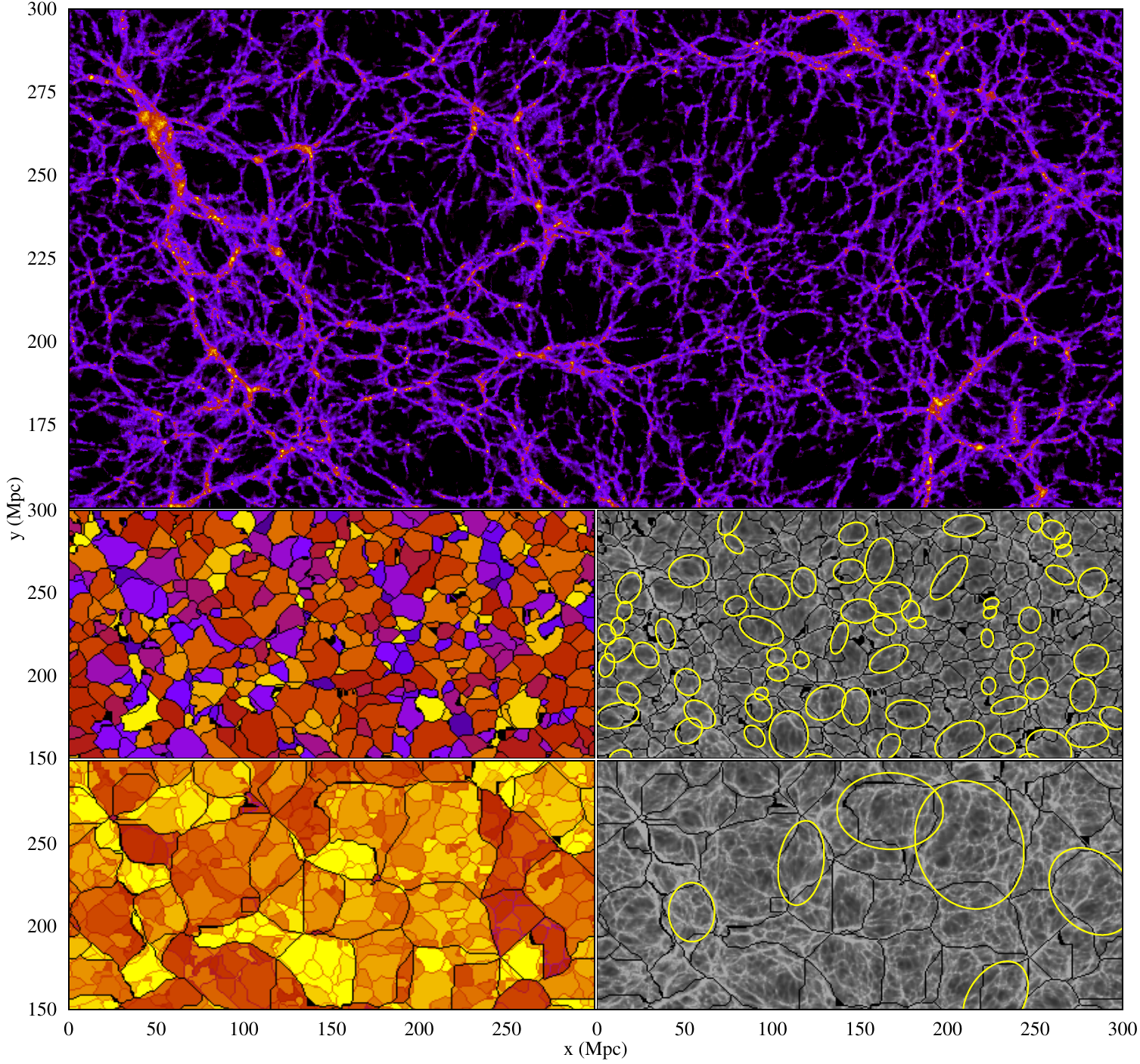}
\caption{\textit{Top}: a density field slice of thickness $0.25\hmpc$ from the $\Lambda$CDM simulation. This slice of 300 by 150 $\hmpc$ is the same as those used in the
bottom panels. Colour represents density as in e.g.\ figures~\ref{fig:model-comparison-z} and \ref{fig:model-comparison-all5}.
\textit{Bottom left two}: the corresponding distribution of voids 
(void borders in black lines, the different colours do not represent anything physical). For the top panel we used a Gaussian filter 
radius $R_f = 1.5 \hmpc$, for the bottom panel $R_f = 6.0 \hmpc$ and the $1.5\hmpc$ ones are transparently inset. \textit{Bottom right two}: a 
random selection of ellipsoid fits (yellow) overlaid on the density field (now in grayscale), again at two radii $1.5$ and $6.0\hmpc$.}
\label{fig:watershedWMAP}
\end{figure*}

To obtain an impression of the application of the WVF formalism, for two different 
scales figure~\ref{fig:watershedWMAP} illustrates the resulting watershed segmentation 
of the density field in the $\Lambda$CDM N-body simulation. In the four bottom
panels, the top two panels have been 
smoothed with a Gaussian filter radius $R_f = 1.5 \hmpc$, the bottom two panels with 
$R_f = 6.0 \hmpc$. In both cases the void boundaries follow the filamentary structures 
quite accurately and the irregular shapes of the voids are well preserved. Also, comparison 
between the watershed void boundaries of the $R_f=1.5\hmpc$ and $R_f=6.0\hmpc$ filtered 
fields shows the loss of void substructure on the larger scale. 

A related tessellation-based method for void identification, ZOBOV \citep{neyrinck08}, 
does yield similar results as WVF \citep{colberg08}. It demonstrates the successful 
application of tessellation-based techniques to identify structures within the cosmic 
matter distribution.

\subsection{Void shapes}
\label{sec:voidProperties}
After determining the void distribution, we calculate properties of each of the voids. This includes some fits of shape 
parameters. In what follows we represent the density field on uniform grid covering the volume of interest. A void 
consists of the group of grid cells that WVF has identified as belonging to a particular watershed basin. 
The geometric centre of a void is defined as the (volume) average of the void's grid cell, or voxel, positions.
The volume of the voids is determined by simply adding the volumes of the voxels that define the void. 

To determine the shape of a void, we fit its volume by an ellipsoid. We assume the density of the void to be uniform.  
The approximation of voids as a homogeneous ellipsoidal region is a more than adequate first order approximation for the interior of 
voids in a wide range of cosmologies, justified by our understanding of the formation of voids around minima in the primordial 
density field \citep[e.g.][]{icke84,weygaert93,sheth04,shandarin06}. To some extent, it is a considerably better description 
for voids than it is for overdense regions. Overdense regions contract into more compact and, hence, steeper density peaks, 
so that the area in which the ellipsoidal model represents a reasonable approximation will continuously shrink. 
By contrast, the region where the approximation by a homogeneous ellipsoid is valid grows along with the void's expansion. 
While voids expand, their interior gets drained of matter and develops a flat ``bucket-shaped'' density profile \cite{weygaert93,sheth04}. 
Hence, the void's natural tendency is to evolve into expanding regions of a nearly uniform density. The approximation is restricted 
to the interior and fails at the void's outer fringes because of its neglect of the domineering role of surrounding material, 
such as the sweeping up of matter and the encounter with neighbouring features. 

The homogeneous ellipsoidal shape allows us to focus entirely on the geometrical properties of the void and avoid possible complications 
introduced by the overdense regions in and around the void. In practice, we proceed as follows. We first calculate the void's 
inertia tensor $I_{ij}$:
\begin{equation}
I_{ij} = \sum_k\left( \delta_{ij} \vec{x}_k^2 - x_{ki} x_{kj} \right) \,,
\end{equation}
where we sum over all cells $k$ belonging to the void. In the above expression, $\vec{x}_k=(x_{k1},x_{k2},x_{k3})$ is 
the distance vector of the $k$-th void cell to the void's centre, and $\delta_{ij}$ is the Kronecker delta. 

The ellipsoidal fit to the void is taken to be the one that would have the same inertia tensor $I_{ij}$ as the void. 

The size, shape and orientation of the ellipsoid are inferred from the eigenvalues and the eigenvectors 
of the inertia tensor $I_{ij}$. The relevant properties in this work are the ellipsoid's semi-axes $a$, $b$ and $c$
($a \geq b \geq c$), which are connected to the eigenvalues of the inertia tensor as:
\begin{align}
a^2 = \frac{5}{2}\left(I_2 + I_3 - I_1\right) \,,\\
b^2 = \frac{5}{2}\left(I_3 + I_1 - I_2\right) \,,\\
c^2 = \frac{5}{2}\left(I_1 + I_2 - I_3\right) \,.
\end{align}
Subsequently, we can characterise the shape of the void by three parameters: the ellipticity\footnote{For consistency, we 
follow \citet{parklee07} and others in using the term ellipticity for this quantity. This term may not be the most adequate, 
as its semantics imply a general description of the shape of an ellipsoid, whereas the total shape of an ellipsoid needs at 
least two parameters. ``Asphericity'' may therefore have been a better term.} $\epsilon$ or conversely the sphericity $s$, the 
oblateness (flattening) $p$ and the prolateness $q$. These quantities are defined as:
\begin{equation}
\label{eqn:shape_parameters}
\epsilon = 1 - \frac{c}{a}\mathrm{,} \quad s = \frac{c}{a}\mathrm{,} \quad p = \frac{b}{a} \quad \mathrm{and} \quad q = \frac{c}{b} \,.
\end{equation}
The ellipsoid's volume is $V = \frac{4}{3} \pi r^3$, where $r = \sqrt{abc}$ is the void's effective radius. Note that this volume 
is slightly different from the real void volume due to irregularity of the void's borders.

To obtain an idea of the resulting void ellipsoids, the bottom righthand panels of figure~\ref{fig:watershedWMAP} depict a random selection 
of void ellipsoids, superimposed on the density field. The top frame shows the ellipsoids corresponding to the WVF voids obtained after 
smoothing the density field by $1.5\hmpc$, while the bottom frame shows the void ellipsoids corresponding to the $6 \hmpc$ smoothed 
density field. In general, there is a good correspondence between the size and shape of the ellipsoids and the voids in the underlying 
density field. Evidently, the fits are rarely perfect as the void boundaries tend to have a rather irregular shape. One consequence of this 
is that the ellipse volumes differ slightly, by an average factor of $\sim 1.08$, from the actual void volumes, determined by adding 
the volumes of the grid cells that WVF identified as belonging to its interior. 

\subsection{Systematics}
\label{sec:error}
The measurement of void shapes in the observational reality, and the corresponding statistical analysis, is subject to a 
number of artefacts and systematic effects. The three main effects that we include in our analysis are galaxy biasing, 
redshift distortions, and cosmic variance. 

\subsubsection{Discrete and diluted samples}
\label{sec:biasing}
By far the most complex effect is that of galaxy biasing. Within the context of this study, we concentrate 
on the consequences of the discrete and diluted nature of galaxy samples with respect to the underlying 
mass distribution, and the biasing of the halo populations. 

We will not address the effects of the far more complex issue of biasing on the basis of 
intrinsic galaxy properties, and refer to an upcoming study in which we systematically address the 
properties of voids in galaxy populations modelled by different semi-analytical models of galaxy 
formation. Here we address the effects of the discrete and diluted nature of the galaxy distribution 
by means of random subsets of simulation particles of the N-body simulations (\S\ref{sec:dm_particles}), 
in combination with samples of haloes from the same simulations. 

\subsubsection{Redshift distortions}
\label{sec:zspace}
The main distorting effect that we consider is that resulting from redshift distortions due to the peculiar velocity of 
matter, haloes and/or galaxies. Given that galaxy redshift survey data distances are purely based on redshifts, 
which inevitably introduces an error due to the considerable contribution of radial velocities to the redshift. 

As a result, the shapes of voids are expected to get systematically distorted. 
The void will apparently expand and, hence, 
in redshift space they will appear more elongated along the radial direction. This will result in a 
redshift space void distribution that is expected to be systematically shifted to slightly higher values of 
the ellipticity $\epsilon$. Also, the orientation of the voids in redshift space will be somewhat anisotropic, with 
a slight tendency to be directed along the radial direction. The effect is rather small, and only becomes 
apparent when considering a large and statistically representative sample of voids \citep[for a thorough recent analytical 
treatment, see][]{shoji12}. 

In this study we restrict ourselves to estimating the (cosmic) variance in ellipticity due to the fact that we
do not know the true direction and magnitude of the peculiar velocity of dark matter and haloes.
We will 
take this into account as an extra error term when assessing the significance of systematic differences between 
the void shape distributions in different cosmologies. To get these estimates, for each of the five cosmologies we 
generate eight redshift space realisations of the matter and halo distributions. Each of the redshift space 
realisations is defined with respect to a randomly chosen location in the simulation box, the ``observer'' location. 
Different observers see different redshift spaces due to the differences in radial velocities with respect to the 
observer. In each of the redshift space realisations we identify the void population and determine the void shape 
distribution using the same tools as in the underlying real space realisation. 

\begin{figure*}
\includegraphics[width=\textwidth]{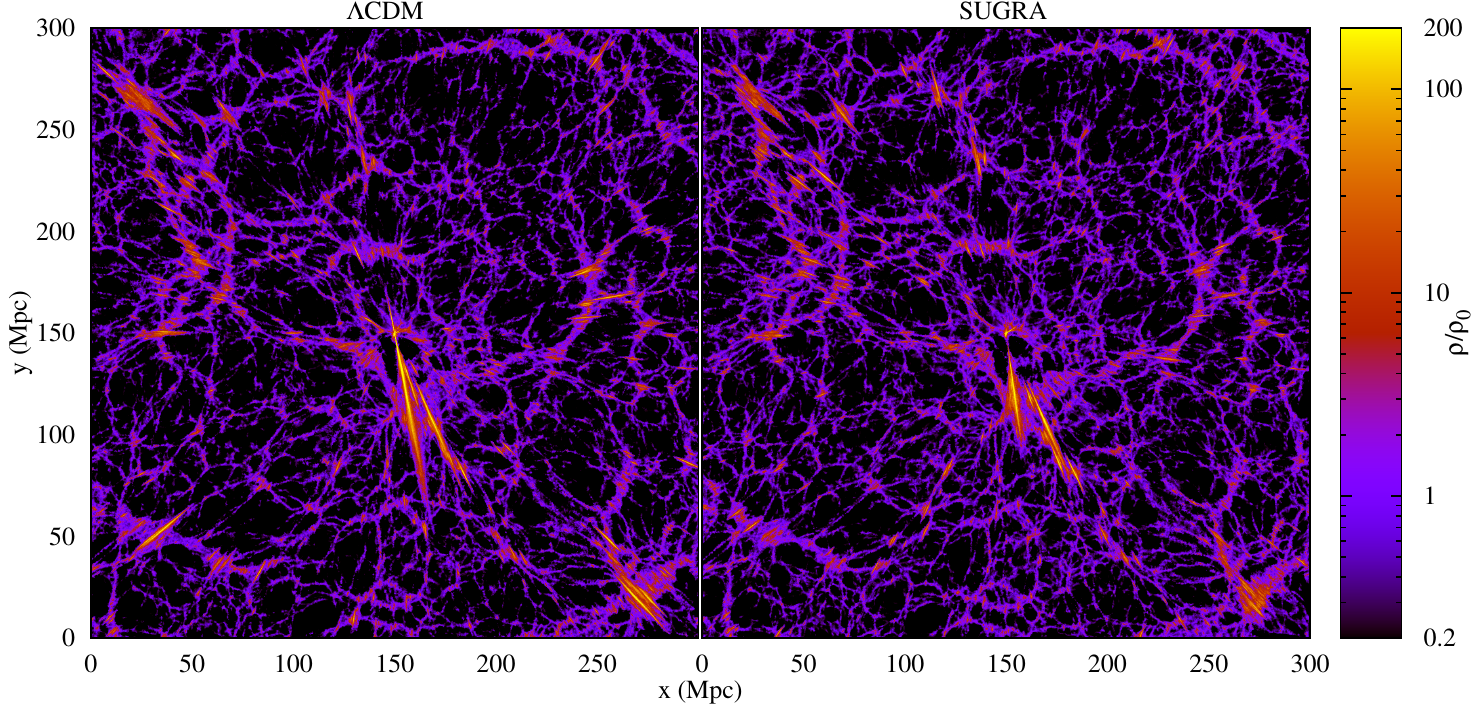}
\caption{Redshift space comparison. Shown are ``redshift space'' (see \S\ref{sec:zspace}) density slices of thickness $0.25\hmpc$ of the simulation boxes of the $\Lambda$CDM (left) and SUGRA (right) models at $z=0$. The observer is placed at the centre. Density values are related to the colours by the bar on the right.}
\label{fig:model-comparison-z}
\end{figure*}

Following the choice of an observer's position, the redshift space realisation is computed from a given 
simulation by transforming the particle locations to \emph{redshift space} with respect to 
the ``observer''. Locating the observer at the centre of the box (and translating the particle and 
halo positions accordingly), the measured redshift of the particle (or halo) is 
\begin{equation}
z = (1+z_\mathrm{H})(1+z_r) -1 \simeq z_\mathrm{H}+z_r \simeq \frac{rH}{c} + \frac{v_r}{c} \,,
\end{equation}
where $v_r$ is the radial component of its peculiar velocity with respect to 
the observer, 
\begin{equation}
v_r = \vec{v}_\mathrm{pec} \cdot \hat{r}\,.
\end{equation}
In this expression, $z_H$ is the cosmological redshift of the particle or halo due to the Hubble expansion. 
Its resulting redshift space position is then given by $(r_\mathrm{z-space},\theta,\phi)$, with $\theta$ and $\phi$ 
its sky position as seen from the observer and  
\begin{equation}
r_\mathrm{z-space} = zc/H \,.
\end{equation}

Examples of resulting redshift space realisations are shown in figure~\ref{fig:model-comparison-z}. They show the 
redshift space structure for a $\Lambda$CDM and a SUGRA cosmology, for the same observer's position and time. Large 
clusters are transformed into the well known \emph{Fingers of God} \citep{jackson72}, whereas voids in redshift space appear to be somewhat 
larger and emptier than in the corresponding real space distribution. Differences between the two models are similar to those seen 
in real space. The most prominent difference is that of the resulting Fingers of God. In the SUGRA cosmology these 
clusters appear to be significantly shorter than their equivalents in the $\Lambda$CDM cosmology. This indicates  
a lower internal velocity dispersion of clusters in the SUGRA cosmology, a manifestation of the slower evolution and 
lower mass of clusters (see \S\ref{sec:simqual}).

\subsubsection{Cosmic variance}
We additionally include the error introduced by cosmic variance.
We calculate the amount of variation in the evolution of the main void ellipticity by taking 
eight different random phase realisations of our low resolution simulations and calculating the standard 
deviations in the eight obtained mean ellipticity values at all redshifts. 

The value of the deviation is quadratically added to the estimated redshift distortion. These combine to the square 
of the total estimated ``error'', which we use in the interpretation of the results presented in the figures of 
\S\ref{sec:shapeVSz}.

\subsection{Analysis setup}
\label{sec:analysis_setup}
In our analysis, we compare the void shape results from the different cosmological samples to each other and to previous results 
from the literature. The analysis involves the following steps:

\begin{itemize}
\item We first compare \textit{our results} to those found in the \textit{literature}. The high 
resolution ($768^3$) simulations are used to see whether the mean void ellipticity indeed 
evolves as predicted, and increases with decreasing redshift. 

\item On the basis of the low resolution ($256^3$) data, we assess whether we can distinguish 
between the dark energy models by comparing their redshift evolution of the mean void 
ellipticity, $\langle\epsilon\rangle(z)$.

\item We compare the \textit{low} and \textit{high resolution} results to check for effects of 
mass resolution. There is no physical reason for there to be a resolution effect, so any effect 
will be due to the methods used.

\item For each of the simulations, we compare the results of the average void shapes obtained for the 
pure \textit{DM particle} distributions with those obtained from the corresponding \textit{DM halo populations}.  
To a large extent the DM halo distribution reflects the galaxy distribution. In terms of estimating 
the effect of its discreteness and sparseness, the DM haloes are largely representative. Results on DM haloes 
may therefore be used to answer the question of whether or not the inferred void shape evolution 
represents a significantly discriminative probe of dark energy in the case of real galaxy redshift 
surveys. 
\end{itemize}
We find marked differences in $\langle\epsilon\rangle(z)$ between DM haloes and DM particles.
To identify the origin of these differences we run the following additional tests:

\begin{figure*}
\includegraphics[width=0.95\textwidth]{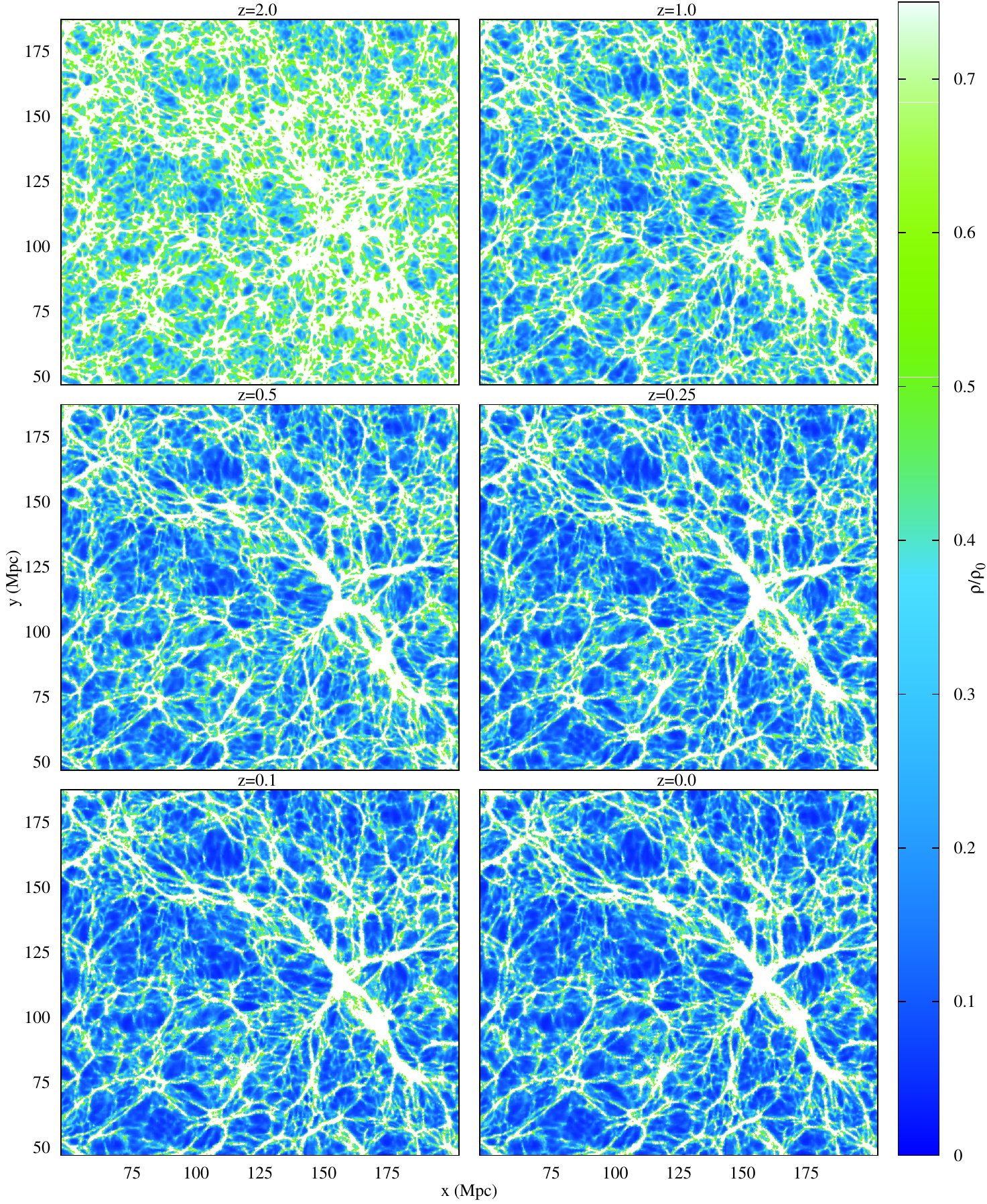}
\caption{Void evolution in $\Lambda$CDM. Part of the density slices (range: $x \simeq [47, 203]$, $y \simeq [47, 188]\hmpc$) of
thickness $0.25\hmpc$ of the simulation boxes of the $\Lambda$CDM model at redshifts (from top to bottom) $z=2.0,\,1.0,\,0.5,\,0.25,\,0.1$ 
and $z = 0$ are shown. The colors were chosen to highlight the evolution of voids, especially in terms of their substructure: 
dark blue indicates the lowest density regions, green and white indicate the higher density regions (see colour bar on the right).}
\label{fig:lcdm-timesteps}
\end{figure*}
\begin{figure*}
\includegraphics[width=0.95\textwidth]{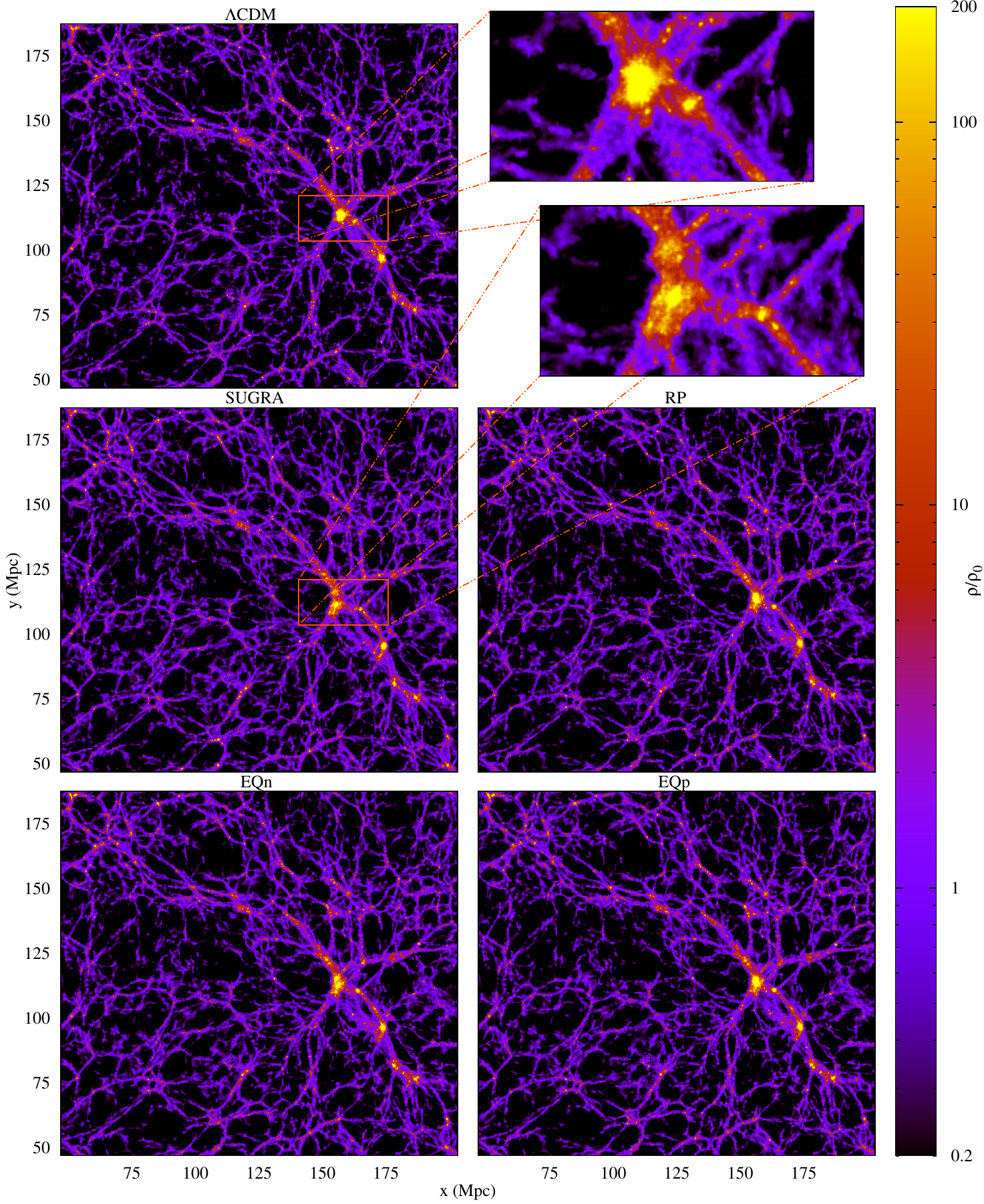}
\caption{Evolution of structure in different cosmologies. Part of the density slices (range: $x \simeq [47, 203]$,
$y \simeq [47, 188]\hmpc$) of thickness $0.25\hmpc$ of the simulation boxes of the $\Lambda$CDM, SUGRA, RP, EQn and EQp
models (from top to bottom) at $z=0$ are shown. Cluster regions are yellow ($\rho/\rho_0 \gtrsim 200$) and voids are black 
($\rho/\rho_0 \lesssim 0.2$), as also indicated in the colour bar to the right. The 
zoom-in boxes (top right) focus on comparable cluster regions in the $\Lambda$CDM and SUGRA simulation. }
\label{fig:model-comparison-all5}
\end{figure*}

\begin{itemize}
\item We compare $\langle\epsilon\rangle(z)$ for the \textit{DM haloes} and samples of the same
number of \emph{randomly sampled} simulation particles from the high resolution simulations. In case the results 
of these two different samples are the same, we may conclude that the void geometry of the halo 
distribution does not intrinsically differ from that of the dark matter distribution.
It would imply that the differences between the void shapes found in the haloes and in the 
dark matter (particle) distribution are mainly the result of the sparsity of the halo population. 
On the other hand, if there are notable differences between the halo voids and the dark matter 
particle voids, the biasing of haloes -- and by implication also those of galaxies -- is an 
important factor. 
\item As an additional -- strong -- test of possible systematic differences between the spatial 
distribution of haloes and dark matter particles, we compare the \textit{random samples} of 
dark matter particles to the \textit{unweighted halo distribution}. The latter consists of the 
same haloes as in the regular halo sample. However, instead of including their true masses, 
we assign the same mass to each halo. If the mean void size $\langle\epsilon\rangle(z)$ differs 
even more in this test than in the previous one, the void population in the halo distribution 
has to be significantly different from that in the dark distribution. This would imply the 
inescapable conclusion that biasing is an important cause for the differences in $\langle\epsilon\rangle(z)$.
\end{itemize}

\section{Results: void population characteristics}
\label{sec:analysis_void}
In this section we will first present some qualitative results on the evolution of the 
WVF identified void population and discuss some overall quantitative characteristics of the void populations.

\subsection{Voids in the dark matter field: visual impression}
\label{sec:simqual}

The evolving dark matter distribution in the $\Lambda$CDM cosmology is shown in figure~\ref{fig:lcdm-timesteps}.
We can see the evolution of several voids (dark blueish regions). We roughly define these as regions with a density contrast
of $\rho/\rho_0 < 0.2$.

Overall, we observe that the most prominent evolution of the void population takes place between redshifts 2 and 0.5. 
This agrees with \citet{huterer01}, who argued that the redshift range $0.2 \lesssim z \lesssim2$ is the most promising 
for probing $w(z)$ \citep[see also][]{hellwing10}. The voids quickly grow from initial seeds with radii of about $1 \hmpc$, 
to voids ranging in size from $2$ to $30 \hmpc$. We also see that in this process the number of voids decreases due to 
the merging of voids, a manifestation of the hierarchical evolution of voids \citep{sheth04,aragon12}. At the later 
time-steps, we may still discern a lot of substructure in voids. These are the remnants of the same structural evolution: 
even billions of years after the voids merge, their outline remains visible as tenuous underdense features in the realm of the 
emerging void \citep[see][]{sheth04}. This is clearly visible in the bottom panels, corresponding to $z=0.1$ and $z=0.0$, 
where we may clearly discern walls in the larger voids. 

In figure~\ref{fig:wvsz} we compared the evolution of the void population in the $\Lambda$CDM simulations with that 
in the other four cosmological models. The most outstanding differences are those between the $\Lambda$CDM cosmology 
on the one hand, and the SUGRA cosmology at the other extreme. 

The large scale structure in the SUGRA cosmology at $z=0$ is less evolved than that in the $\Lambda$CDM cosmology. 
One of the manifestations of this concerns the voids, whose size is smaller. Also, the filaments in the SUGRA cosmology 
are more diffuse, and clusters are less pronounced and clumpier. The lower level of evolution in the SUGRA model ties 
in with the higher value of $w$ over a large fraction of its cosmic evolution. The closer the value of the dark energy 
equation of state $w$ is to its upper limit of $-1/3$, the shorter the cosmic timescale over which structure can evolve. 
The SUGRA universe, for instance, is effectively $\sim560$ Myr younger at $z=0$ than the $\Lambda$CDM universe.
One can also observe this from the measured values of $\sigma_8$. As shown in figure~\ref{fig:sigma8vsz}, at each 
redshift SUGRA has a lower value of $\sigma_8$ than that found in the other cosmologies, while $\Lambda$CDM has the 
largest amplitude of mass fluctuations and structure. 

The RP, EQn and EQp dark energy models differ less dramatically from the $\Lambda$CDM model. All largely follow the same 
evolution, and it is more difficult to visually distinguish the resulting structures from those seen in the $\Lambda$CDM cosmology
than it was for the SUGRA cosmology. 
The void regions observed in the corresponding frames of figure~\ref{fig:model-comparison-all5} have a large degree of similarity 
in shape and size to those found in the $\Lambda$CDM simulation.  

To emphasize the large difference between the $\Lambda$CDM cosmology on the one hand, and the SUGRA cosmology on the 
other hand, we zoom in on a central cluster region (see zoom-ins in figure~\ref{fig:model-comparison-all5}). When 
comparing these objects, we should note that the corresponding simulations started from the same initial conditions, 
so that any difference in structure, morphology and dynamics is a direct reflection of the influence of dark energy 
on the structure formation process. Not only do we find that the $\Lambda$CDM cluster is more compact and centrally 
concentrated, but also that the morphology of the surrounding matter distribution is substantially different. The 
weblike filamentary structures in the SUGRA cosmology find themselves in a dynamically younger state. The number 
of filaments is higher, while they are thinner and less well defined. In the $\Lambda$CDM situation, we find the 
central cluster embedded in a web of a few strong and well defined filaments. 

\subsection{Void sizes and shapes}
\label{sec:morph-vol}
Using the void population obtained by means of the WVF, the size and shape of each of the voids is calculated 
following the description in \S\ref{sec:voidProperties}. 

\subsubsection{Different cosmologies at $z=0$}
\begin{figure}
	\subfigure[Radii $r$]{\label{fig:radii-models}\includegraphics[width=\columnwidth]{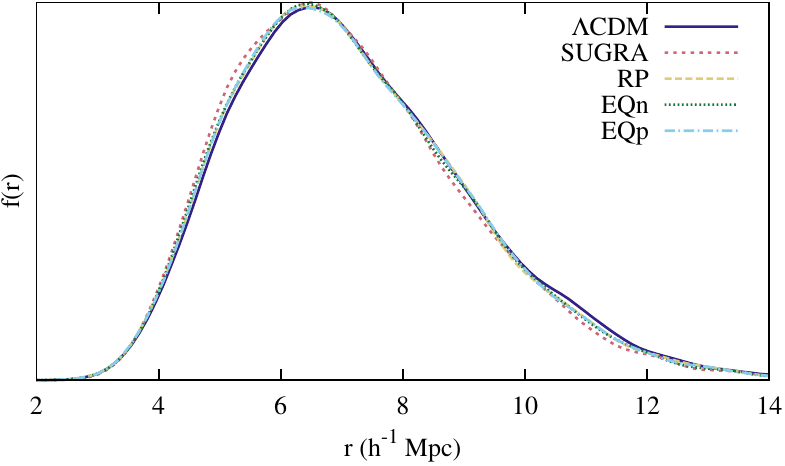}}
	\subfigure[Ellipticities $\epsilon$]{\label{fig:ell-models}\includegraphics[width=\columnwidth]{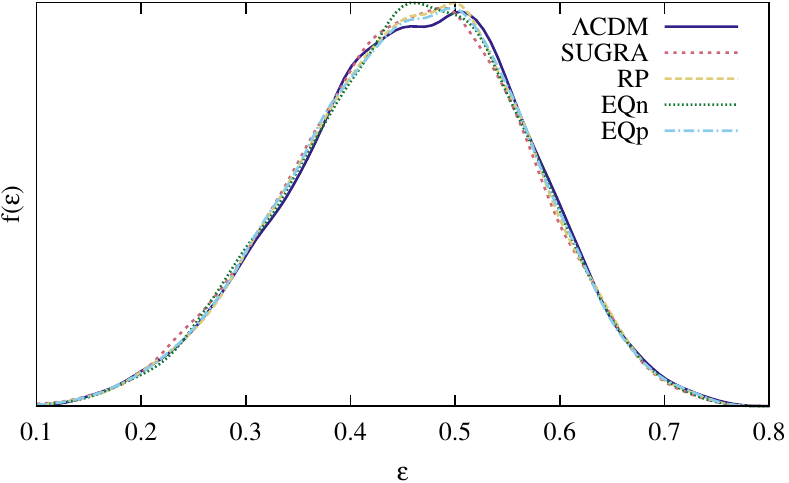}}
	\caption{Void radius and ellipticity PDFs ($R_f = 1.5 \hmpc$) in the different dark energy model simulations at $z=0$.}
	\label{fig:shapes-models}
\end{figure}

In figure~\ref{fig:shapes-models} we plot the (effective) void radii 
$r$ and ellipticities $\epsilon$ in the five cosmologies, at $z=0$. These 
are all computed from the high resolution simulations. The functions shown in the plots are the normalised 
probability density functions, for which
\begin{equation}
\int_{-\infty}^{\infty} f(r)dr=1\,; \quad \int_{-\infty}^{\infty} f(\epsilon) d\epsilon =1\,.
\end{equation}
These results were obtained at a smoothing radius of $R_f=1.5\hmpc$.

One conclusion from this is that the effects we are trying to measure are rather subtle. To 
uncover systematic trends, we therefore will have to direct our attention to moments of the 
ellipticity distribution and average over the entire sample. In this respect, we follow earlier 
studies in concentrating on the mean ellipticity.

\begin{figure}
	\subfigure[Radii $r$]{\label{fig:radii_filterRadii}\includegraphics[width=\columnwidth]{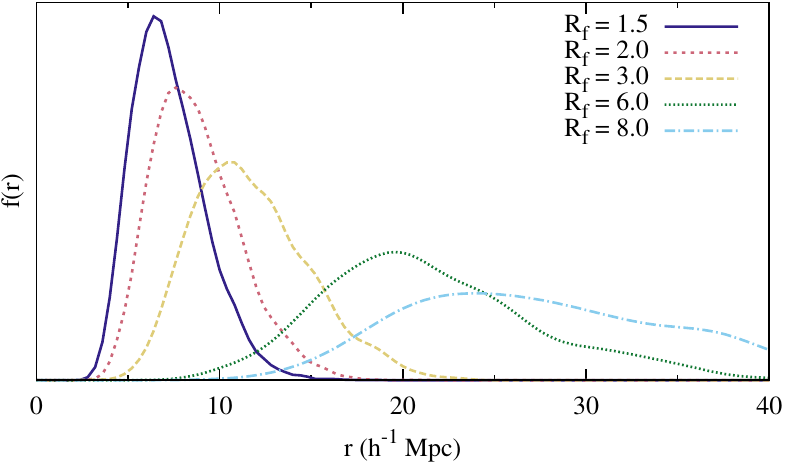}}
	\subfigure[Ellipticities $\epsilon$]{\label{fig:ell_filterRadii}\includegraphics[width=\columnwidth]{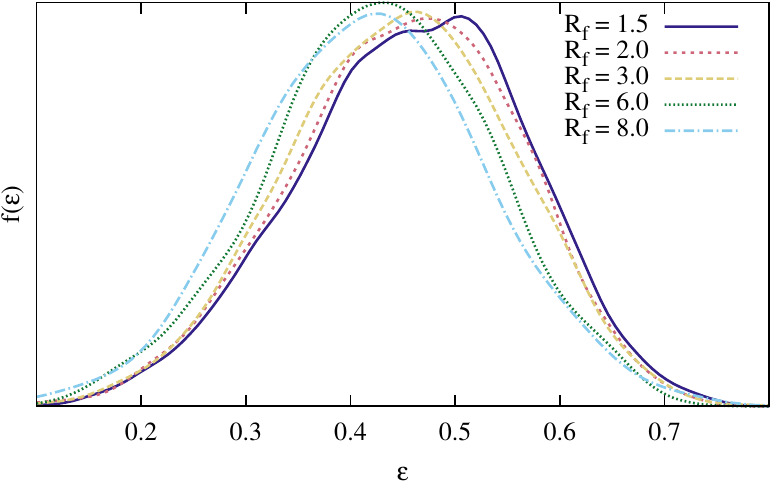}}
	\caption{Void property PDFs for different WVF smoothing scales $R_f$.}
	\label{fig:filterRadii}
\end{figure}

\subsubsection{Scale dependence}
One of the determining factors in our analysis is the scale at which we study the void population. 

The effect of the smoothing radius on the size distribution of the voids is shown in figure~\ref{fig:radii_filterRadii}.
We see that the size of the detected voids increases systematically with smoothing scale. 
This is because the small scale void boundaries are smoothed out, resulting in a sample of larger void regions 
(see figure~\ref{fig:watershedWMAP}). The distribution is peaked around a characteristic void size. The void 
population is marked by a large number of voids around this characteristic void size, with hardly any smaller voids and 
a strongly declining number of larger voids. This distribution agrees entirely with that predicted within the context of 
the excursion-set description of the hierarchical void evolution \citep{sheth04}. As is born out by the figure and as 
expected, the peak size -- as well as the entire distribution -- systematically shift towards higher values as the smoothing 
scale increases.
Indications towards this behaviour have independently also been found in other studies. One example is the 
recent study by \citet{einasto11}, who used a different void finder and different void-delineating objects. They noted 
a systematic shift towards higher (mean) void radii with increasing lower threshold mass of the defining objects. This 
interesting parallel was to be expected, as the high mass objects will form preferentially in the high density regions 
that will dominate the cosmic web on the largest (smoothing) scales.

Of more immediate interest for our program is the behaviour of void shapes as a function of scale. 
The void shape distribution turns out to vary only mildly with the scale of the density field, as shown in figure~\ref{fig:ell_filterRadii}.
To a first approximation, the shape distribution of the void population appears to have a nearly scale invariant
character. There is a slight tendency for the voids on the largest scales of 
$6.0$ and $8.0\hmpc$ to be slightly more spherical. However, to a large extent this may be ascribed to the 
fact that it concerns samples with a considerably lower number of objects and hence with a larger statistical 
uncertainty. Nonetheless, the nearly scale independent behaviour of the void shape distribution is an 
interesting and highly relevant result. It implies that -- in the case of the dark matter voids -- we can restrict 
ourselves to an evaluation of the shapes of voids on a scale of $R_f = 1.5 \hmpc$, as this would contain 
all necessary shape information and guarantee a statistically optimal result.

\subsubsection{Overdense void boundaries}
One feature of the WVF algorithm that requires some attention is the fact that, in its pure form, it treats 
the entire interior of each void basin as belonging to a void. The only regions which are not included are 
the boundary grid cells where two voids meet. We investigated whether the inclusion of \textit{overdense regions} 
at the boundaries of detected voids might cause artefacts in the inferred void shape distribution. Only the inner 
parts of the void are properly described by a homogeneous ellipsoidal model, 
\footnote{Because the ``role of surrounding material [will dominate], through the sweeping up of matter and the encounter 
with neighbouring features'' \citep{weygaert11}.} and are therefore expected to adhere better to the ellipsoidal fit of 
our analysis. 

By removing the overdense boundary regions, about 14\% of the volume in voids is eliminated. The shape analysis based 
on these voids does indeed yield slightly different results. One finding is that the mean ellipticities tend to 
be slightly higher. However, the offset is quite comparable for all simulations, at all redshifts. There are no changes 
with respect to any of the qualitative results discussed in the following sections. We therefore decided to keep to the 
basic WVF algorithm.

\subsection{Voids in redshift space}

\begin{figure}
	\includegraphics[width=\columnwidth]{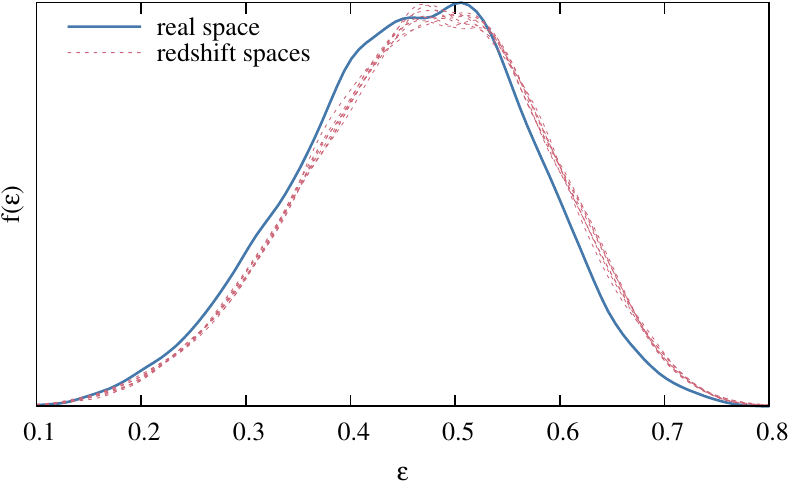}
	\caption{Void ellipticity PDFs ($R_f = 1.5 \hmpc$) in the $\Lambda$CDM model in real space and in 8 different redshift spaces.}
	\label{fig:ell_zspaces}
\end{figure}

In figure~\ref{fig:ell_zspaces} we compare the ellipticities of voids in the $\Lambda$CDM dark matter distribution with 
that in eight redshift space realisations within the same high resolution simulation. We find a rather consistent offset 
of the ellipticity distribution of around $\Delta \epsilon \sim 0.014$. The redshift space voids have, as expected, 
a systematically more elongated shape. In the other cosmologies, the situation is comparable, but the effect is somewhat smaller. 

In addition to the systematic shift, which represents the systematic redshift space distortion, there is also a sizeable 
scatter between the different redshift space realisations. This is the error which is introduced by local variations 
in the contribution of peculiar velocities to the redshift of objects. It behaves like an extra contribution to 
the cosmic variance. The scatter is also a function of filter scale $R_f$, varying from 0.00076 in the case of 
voids on a scale of $R_f=1.5\hmpc$ to 0.0012 for voids on a larger scale of $R_f=3.0\hmpc$. In the subsequent 
analysis we restrict ourselves to voids on these scales. 

\begin{figure}
	\subfigure[Radii $r$]{\label{fig:radii-haloes-scales}\includegraphics[width=\columnwidth]{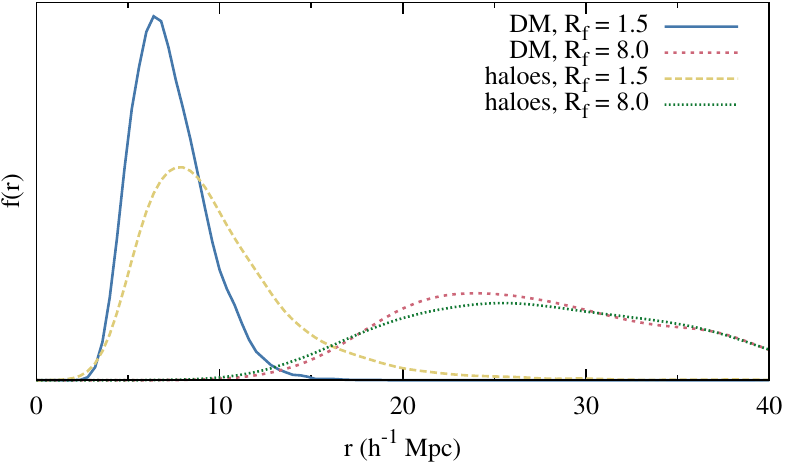}}
	\subfigure[Ellipticities $\epsilon$]{\label{fig:ell-haloes-scales}\includegraphics[width=\columnwidth]{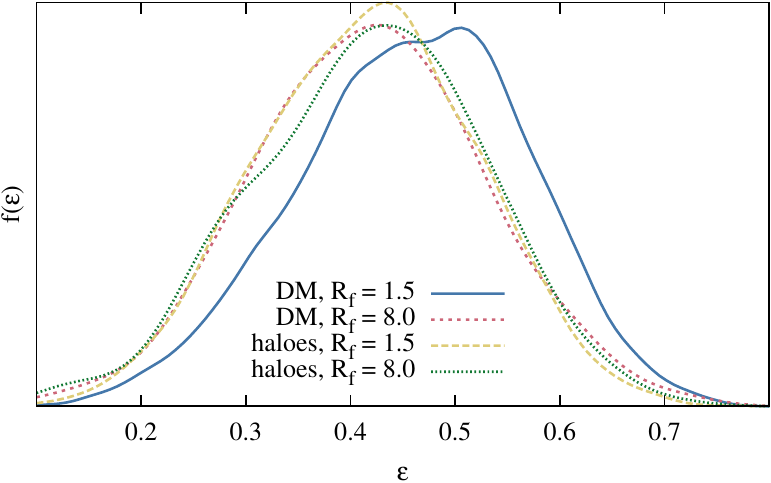}}
	\caption{Void radius and ellipticity PDFs of voids in the dark matter and halo distributions, at different scales.}
	\label{fig:shapes_haloes_scales}
\end{figure}

\subsection{Voids in the halo distribution}
When comparing the spatial halo distribution in figure~\ref{fig:haloslice} with the underlying dark matter distribution, we 
observed the substantial loss in spatial resolution. The halo distribution, even at $z=0$, is much sparser than the 
dark matter particle distributions in the simulations. The diluted halo sample may therefore not be expected to trace 
the fine features visible in the matter distribution, and certainly will not be able to accurately sample the 
small scale void population. 

In how far this affects our study may be appreciated from figure~\ref{fig:shapes_haloes_scales}. At a scale 
of $R_f=1.5\hmpc$, the haloes completely fail to reproduce the size distribution of the voids in the underlying 
void distribution (figure~\ref{fig:radii-haloes-scales}). This may be hardly surprising: haloes cannot 
resolve these structural features. Along with this, we also see that the halo distribution must be used with care when 
investigating the void ellipticity distribution (figure~\ref{fig:ell-haloes-scales}). 

On the other hand, at a large scale, like $R_f=8 \hmpc$, we find complete accordance between the void population 
in the dark matter distribution and the voids found in the halo distribution. Both their size and shape 
distribution are, within acceptable limits, similar. 

\subsubsection{Scale dependence and sample size}
In the generic observational situation, where the voids are traced by a discrete and dilute galaxy distribution, 
the scale dependence of the void analysis involves two opposing effects. On the one hand, it is easier to reliably 
trace the outline and measure the shape of larger scale voids. On the other hand, the number of voids traced by the 
object sample within a given volume of observed space will strongly decrease as we consider voids on larger scales. The 
errors in the measured statistical moments will increase accordingly.

Hence, it is necessary to find a compromise between both requirements. However, the sampling density 
of objects may get so sparse that it becomes unfeasible to trace enough voids on a sufficiently large 
scale. In that situation, it will be impossible to use the void population for an attempt to measure 
the dark energy equation of state as any such measurement will suffer from inherently large errors.

One possible improvement of this will be to substantially increase the survey volume. This would 
enable to obtain more reliable and representative statistics for the larger number of 
small scale voids. 

\begin{figure}
	\includegraphics[width=\columnwidth]{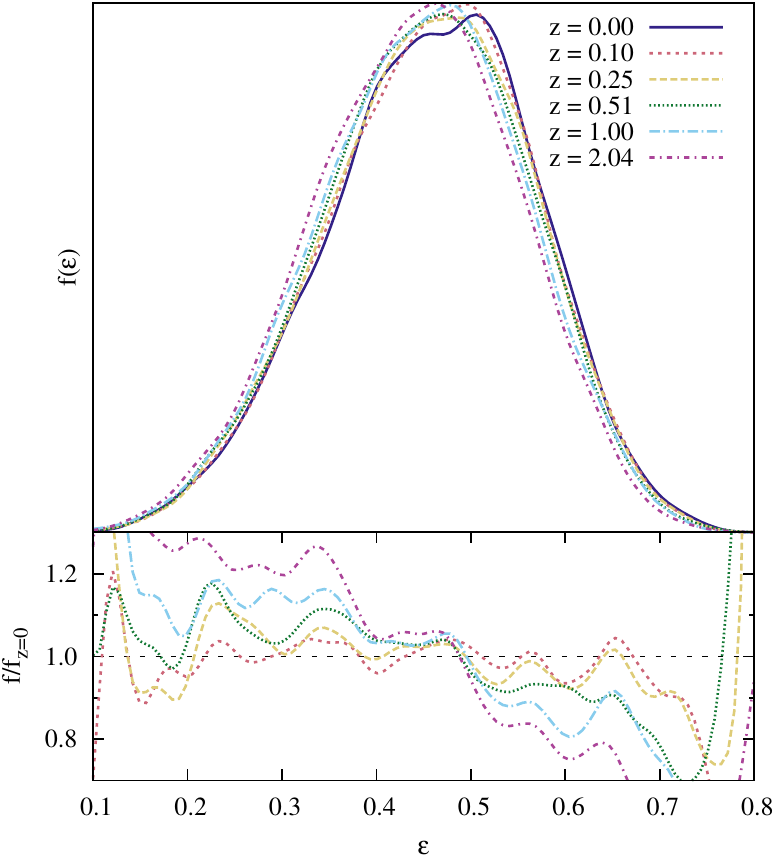}
	\caption{Void ellipticity PDFs ($R_f = 1.5 \hmpc$) in the $\Lambda$CDM model simulations at different redshifts. The bottom panel shows 
the ratio of the PDFs at $z\neq0$ to that at $z=0$.}
	\label{fig:ell_WMAPevol}
\end{figure}

\section{Results: void shape evolution}
\label{sec:shapeVSz}
Having established the basic size and shape properties of the void population in the different simulated 
cosmologies, we arrive at the examination of the time evolution of the mean ellipticity, 
\begin{equation}
\langle \epsilon \rangle(z)\,=\,\int d\epsilon \, \epsilon f(\epsilon,z)\,
\end{equation}
and its relation with the character of the dark energy in the corresponding cosmology. 

\subsection{Basic results vs literature}
\label{sec:bosVSliterature}
The void ellipticity distribution in the $\Lambda$CDM cosmology evolves systematically with redshift. This is 
clearly visible in figure~\ref{fig:ell_WMAPevol}, where we find a gradual shift of the ellipticity distribution 
towards higher values of $\epsilon$ as time proceeds. In other words, the ellipticity of voids is expected 
to decrease towards higher redshifts. This is entirely in line with the expected generic behaviour, as described by 
e.g. \citet{leepark09}. 

The evolutionary trend of the mean ellipticity $\langle\epsilon\rangle$ follows the general trend of \citet{leepark09}, 
as is clearly shown in figure~\ref{fig:evsz_highVsLow}. The mean ellipticity of the voids increases with time, 
with voids being less elongated towards higher redshifts. In one respect our results differ with those obtained by 
\citep{leepark09,lavaux10}: the voids in our simulations do not have equal ellipticities at $z=0$. This is a result 
of the difference in normalisation between our simulations. By normalising our simulations, via $\sigma_8$, at 
the recombination redshift $z=1089$, the level of structure formation at $z=0$ between the different cosmologies  
may be expected differ. The differences in mean void shapes is one particular manifestation. 

When comparing the $\Lambda$CDM and SUGRA results for the ellipticity evolution, we find that there is 
a significant redshift range over which we can clearly distinguish between the void ellipticities in the 
different cosmologies. 

\subsection{Simulation resolution}
\label{sec:lowVShigh}

\begin{figure*}
 \subfigure[Mean ellipticity as a function of redshift. $\Lambda$CDM high and low resolution simulations and the SUGRA low resolution one.]{
  \label{fig:evsz_highVsLow}
  \includegraphics[width=\columnwidth]{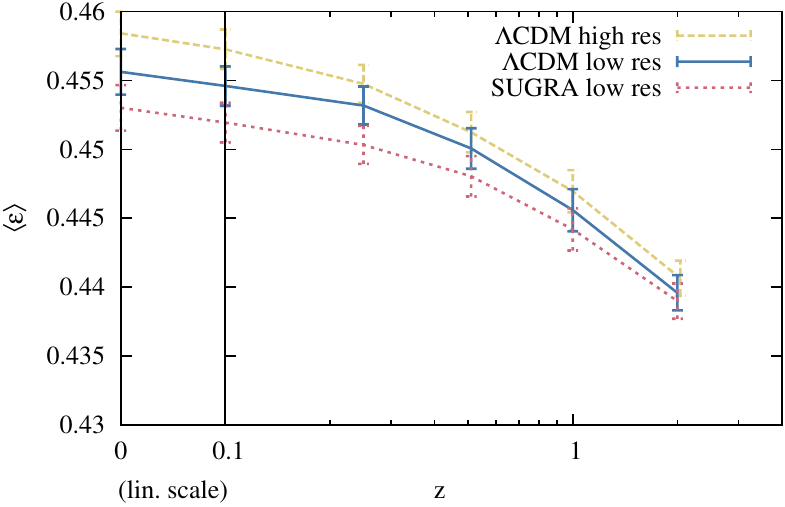}
 }
 \hfill
 \subfigure[Mean ellipticity as a function of redshift. $\Lambda$CDM and SUGRA low resolution DM particle samples and the DM halo samples.]{
  \label{fig:evsz_DMvsHaloes}
  \includegraphics[width=\columnwidth]{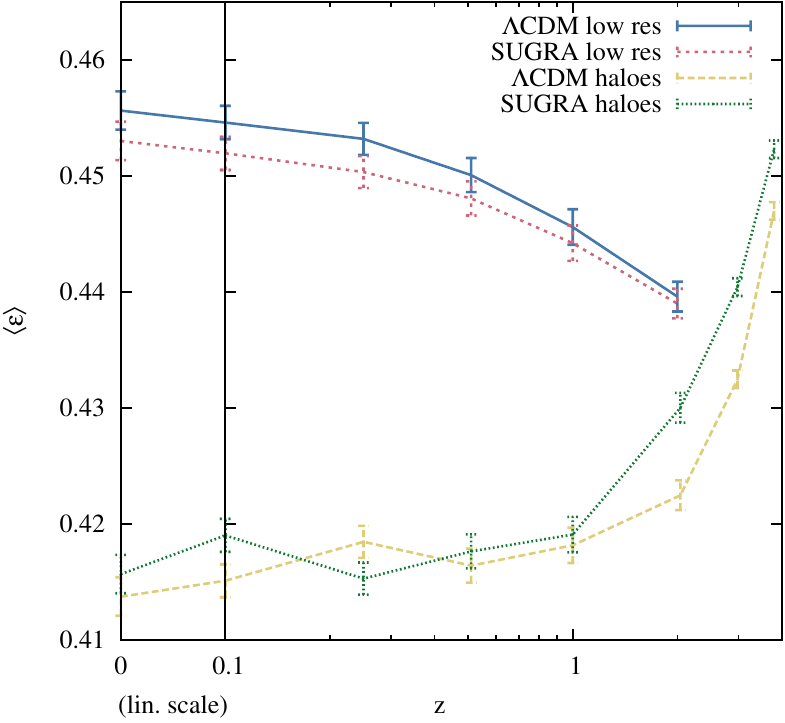}
 }
 \subfigure[Mean ellipticity as a function of redshift. $\Lambda$CDM weighted halo population and a random DM particle sample with the same number of particles as the halo set. Results for WVF filter radii of $1.5$ and $3.0\hmpc$ are shown.]{
  \label{fig:evsz_subsetVsWeighted}
  \includegraphics[width=\columnwidth]{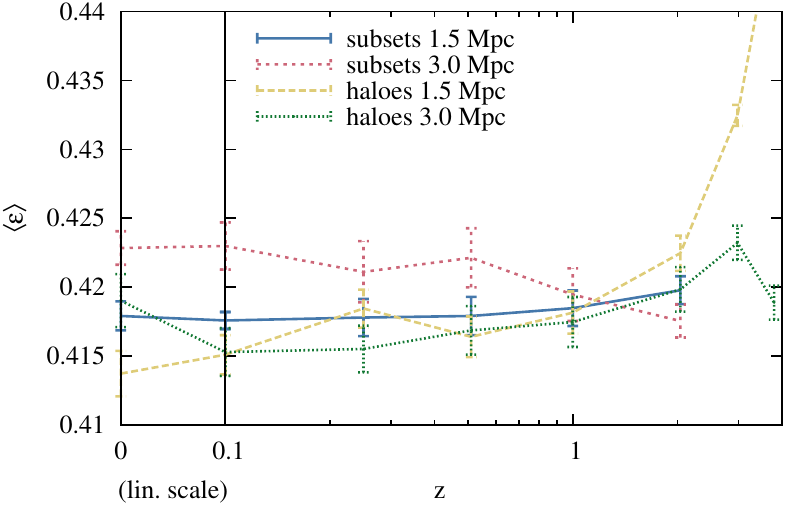}
 }
 \hfill
 \subfigure[Mean ellipticity as a function of redshift. Similar to figure~\ref{fig:evsz_subsetVsWeighted}, but with unweighted halo population results.]{
  \label{fig:evsz_subsetVsUnweighted}
  \includegraphics[width=\columnwidth]{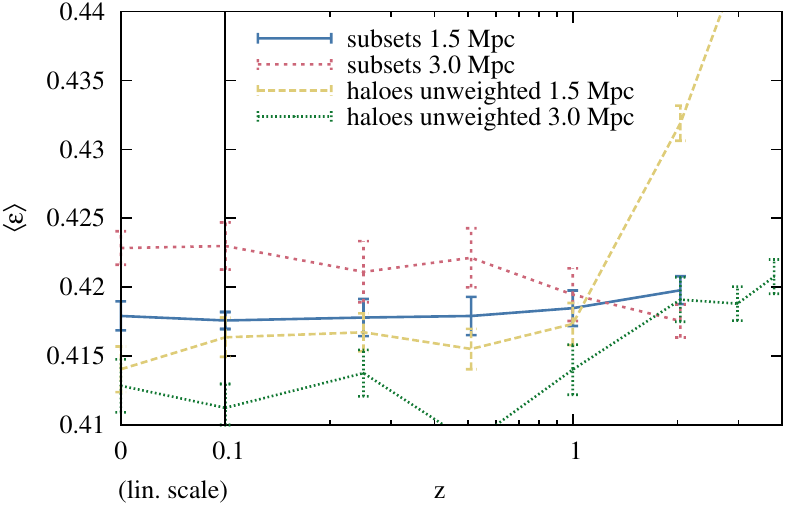}
 }
 \subfigure[Mean ellipticity as a function of redshift. The low and high resolution $\Lambda$CDM results are shown at higher filter radii. The low and high res simulations converge at these higher filter radii, because the small scale differences are smoothed out.]{
  \label{fig:filter_radius_convergence}
  \includegraphics[width=\columnwidth]{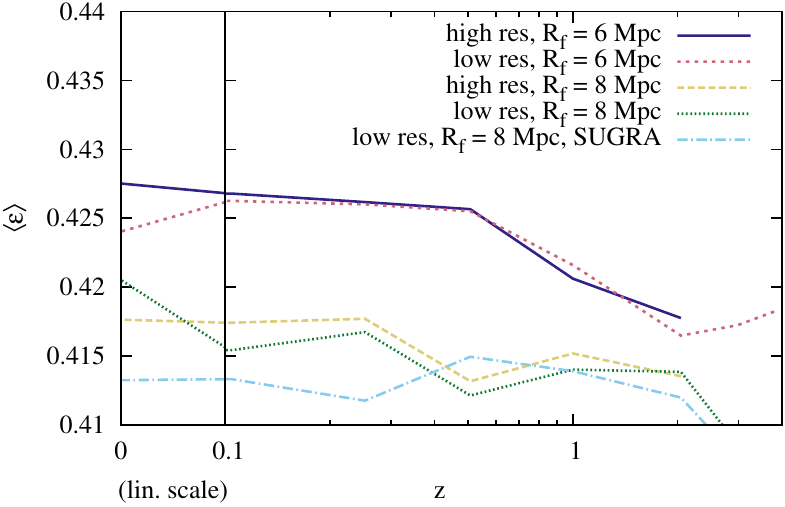}
 }
 \caption{Mean ellipticity as a function of redshift, $\langle\epsilon\rangle(z)$. In the four plots above, in the right part the $z$-axis is in logarithmic scale and the left part is in linear scale (so we can show $z=0$ as well).}
 \label{fig:evsz}
\end{figure*}

When comparing the inferred mean void ellipticity evolution in the low resolution 256$^3$ $\Lambda$CDM simulations and the 
high resolution 768$^3$ $\Lambda$CDM simulations, we find that there is a slight resolution effect. We analysed the two 
simulations, which have exactly the same initial conditions -- in terms of mode amplitudes and phases -- and cosmological 
parameters. Figure~\ref{fig:evsz_highVsLow} reveals the difference between the high resolution (yellow) and 
low resolution (blue) simulation. The only explanation is the difference in mass resolution of the simulations.

To confirm that the difference is purely a resolution effect, we compared the low and high resolution simulations at 
large filter radii $R_f$, to smooth out the small scale differences caused by the different resolutions. Indeed, we 
see in figure~\ref{fig:filter_radius_convergence} that at large scales, the results for the high and low resolution 
simulations converge.

We also find that at these large scales the power of the analysis to discriminate between different dark energy 
cosmologies is lost. The figure reveals that it is not possible to distinguish significantly between 
the ellipticity evolution curves obtained from the SUGRA simulation and the corresponding $\Lambda$CDM curves. 
This strengthens our choice to opt for a study of the void population at filter scales of $R_f=1.5\hmpc$ and 
$R_f=3.0\hmpc$. 

\subsection{Voids in the halo distribution}
\label{sec:dmVShaloes}
Given that the resolution of the DM halo sample, of around 560,000-580,000 particles, is considerably lower than 
that in the low resolution DM particle sample we may not be surprised to find that it is difficult to 
find any significant evolutionary trends in the corresponding void population. 

We indeed find that it is not possible to detect a decrease in mean ellipticity of the void population. Rather,
we find that the halo void ellipticities stay remarkably constant up to at least $z=1$ and 
rise steeply towards higher redshifts (see figure~\ref{fig:evsz_DMvsHaloes}). The latter is probably a result 
of the (strongly) decreasing number of haloes at higher redshifts and the corresponding poorer sampling of the 
underlying full density field. Any hope seems lost to discriminate between different dark energy models on the 
basis of the measured void shape parameters. 

\subsection{Sparsity effect}
\label{sec:haloesVSsubsets}
A key question for understanding the inability of haloes to reproduce the ellipticity evolution of 
voids is whether this is mostly an effect of the discreteness and sparseness of the halo sample or 
whether intrinsic biasing is also at play. 

To this end, we take a random subsample of particles from the dark matter particle distributions and 
repeat the analysis. The number of randomly sampled particles is taken to be equal to the halo population. 
Bootstrapping errors are used to obtain error estimates of the inferred mean void shapes. They are the 
standard deviations of the eight mean ellipticities that were obtained from eight different random subsets 
of the high resolution DM particle set.

Figure~\ref{fig:evsz_subsetVsWeighted} shows that the (lack of) void shape evolution in the diluted random 
DM particle distribution largely agrees with the results of the halo void study. This implies that the 
deviation of the halo void shapes from the shapes of voids in the high resolution dark matter simulations 
is to be ascribed to the sparsity of the halo population. 

However, we also observe some additional differences. On a scale of $R_f=3\hmpc$, the void shape evolution 
in the random subsample appears to differ significantly from that of the halo void ellipticity curves. 
Moreover, the increase of ellipticity with redshift that 
is observed in the halo void sample (at a scale of $R_f=1.5\hmpc$), is not reproduced by the 
voids in the subsampled particle distribution (at $R_f=3\hmpc$). This argues for the influence of additional effects, 
in particular that of the spatial bias of haloes. 

This conclusion is confirmed when considering the results obtained for the void population in the 
unweighted halo distribution (see \S\ref{sec:analysis_setup}). The differences with the 
random subsample voids become slightly larger than in the case of the regular weighted halo 
voids, especially at $R_f = 3 \hmpc$ (see figure~\ref{fig:evsz_subsetVsUnweighted}). 

The inescapable conclusion appears to be that the spatial bias of the halo population, and 
all accompanying complications, is indeed a factor of significant importance when seeking to 
use the void population as tracer of the dark energy equation of state. 

\section{Shapes and clustering: the $\sigma_8$ degeneracy}
\label{sec:sigma8}
We looked into the possibility that the differences between the model ellipticities are not primarily 
caused by differences in the equation of state parameter $w(z)$ of dark energy, but by some other 
cosmological parameter. We have come to the interesting conclusion that the one exclusive and dominant 
factor is that of the level of clustering and structure development, expressed by $\sigma_8$. 
In other words, the void ellipticity distribution is a manifestation of the level of clustering 
of matter.

At the present epoch, the different dark energy models have different values of $\sigma_8$. This is 
the result of the different structure growth rates between the different cosmologies, all starting 
from the same primordial density field whose amplitude has been normalised at the time of last 
scattering. The value of the amplitude is fixed by the value of $\sigma_8$ measured from WMAP data. 

\begin{figure}
\includegraphics[width=\columnwidth]{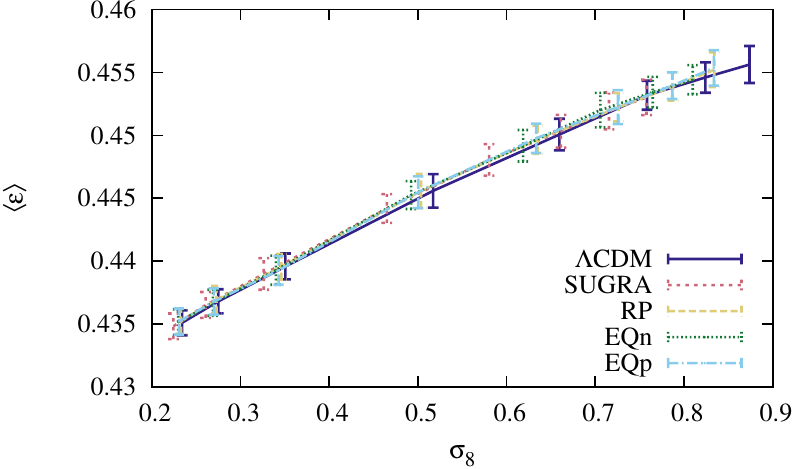}
\caption{Mean ellipticities versus $\sigma_8$. The different lines show the low resolution simulations of the five different cosmologies. The lines consist of $\sigma_8$ values at redshifts $0$, $0.1$, $0.25$, $0.51$, $1.0$ and $2.04$.}
\label{fig:sigma8vsell}
\end{figure}

In figure~\ref{fig:sigma8vsell} we see that there is a strong correlation between $\sigma_8$ and the mean 
ellipticity, independent of dark energy model or redshift. Both lines consist of $\sigma_8$ measurements 
at redshifts of $0$, $0.1$, $0.25$, $0.51$, $1.0$ and $2.04$. If they had been redshift dependent, we should 
have been able to distinguish e.g.\ the point at the middle of the SUGRA line (which corresponds to $z\approx0.5$) 
from the point at the same location of the $\Lambda$CDM line (which corresponds to $z\approx1$). The strong 
correlation between $\sigma_8$ and ellipticity is especially surprising when we realise that the $\Lambda$CDM and 
SUGRA models differ substantially in almost ever other aspect that we investigated. Yet, there is little to no 
difference between these two models in figure~\ref{fig:sigma8vsell}.

It seems then that the main cause of the differences between these models is, in fact, $\sigma_8$. Of course,
in this case, the difference in dark energy models, i.e.\ in $w(z)$, is the underlying cause of the differences
in $\sigma_8$. $w(z)$ exerts its influence on $\sigma_8$ through $D(z)$ (equation~\ref{eqn:sigma8norm}), which
in turn contains a factor $H(z)$, thus realising the coupling to $w(z)$ (equation~\ref{eqn:friedmann}). 

\section{Conclusions and discussion}
\label{sec:discussion}
We have investigated the suggestion that the evolution of the shape of voids is a 
sensitive probe of the nature of dark energy. On the basis of a set of N-body 
simulations of structure formation in five different dark energy cosmologies, 
including dynamical dark energy models, we confirm this sensitivity in the case 
of voids in the dark matter distribution. 

The first observation is that the SUGRA model has a less developed void 
population and shows a lower level of clustering than the $\Lambda$CDM model, 
with the other quintessence models representing intermediate cases. The fact 
that the extended quintessence models do not have a more manifest and pronounced 
signature of their rather distinct nature is one of our surprising findings. 

A key component of our assessment is in how far this finding is affected by the 
sparsity and spatial bias of the objects that probe the underlying dark matter 
distribution. In other words, whether this dark energy sensitivity may be exploited 
when studying the void population in the observed galaxy distribution. We find that 
the discrete, sparse and biased character of the halo and galaxy distributions 
is seriously impeding the potential of probing the nature of dark energy from voids 
traced out by these objects. 

\subsection{The $\sigma_8$ degeneracy}
We have also looked deeper into the relation between the shapes of voids and the 
amplitude of the density fluctuations. Following the relations derived by 
\citet{parklee07}, we know that the ellipticity of voids is largely a reflection 
of $\sigma_8$, which quantifies the average amplitude of density fluctuations. 
In this study, we have shown that most of the effect is indeed degenerate with respect to 
$\sigma_8$. 

In fact, we find that differences in the void shape $\langle\epsilon\rangle(z)$ between the various dark energy models can almost be 
completely ascribed to differences in $\sigma_8(z)$, i.e.\ the amplitude of density fluctuations. Of course, the differences 
in $\sigma_8$ and cosmic growth factor at any one cosmic epoch between the different cosmologies is the result of the 
differences in the equation of state parameter $w(z)$. As is clearly borne out by figure~\ref{fig:sigma8vsell}, the amount 
of clustering fully determines the evolution of the void ellipticity. The fact that $\langle\epsilon\rangle(z)$ and $\sigma_8$ 
are so strongly correlated is interesting by itself. As they encapsulate two different aspects of the large scale mass 
distribution, we would not immediately have suspected the existence of such a strong one-to-one connection.

This leads us to the conclusion that if we wish to use void shape as a probe of dark energy, we need to measure 
not only the void ellipticity $\langle\epsilon\rangle(z)$ but also, independently, $\sigma_8(z)$ at the same redshift. 
This will enable us to break the degeneracy between $\langle\epsilon\rangle(z)$ and $\sigma_8(z)$.
It will be a considerable challenge to improve the accuracy of $\sigma_8(z)$ measurements at all redshifts to anything 
comparable to that determined from the CMB. This also involves a better understanding of the non-linear effects on 
the evolution of $\sigma_8$ \citep{juszkiewicz10a}. Recent measurements of the growth factor $f(\Omega)$ as a function 
of redshift are a great step in the right direction \citep[e.g.][]{blake11, nusser12, tojeiro12, blake12}.

\subsection{Probing dark energy in the observational reality}
The discrete and sparse nature of the halo and galaxy distribution form a major source of confusion in 
retrieving the signal of void ellipticity evolution. This adds to the confusion as a result of redshift 
distortions in the inferred galaxy distribution maps. 

We have tested the void ellipticities in a halo distribution whose average density is around $0.019 h^3 \mathrm{Mpc}^{-3}$, 
comparable to that of the SDSS DR7 galaxy sample \citep{montero09}. Volume limited samples will have even lower sampling 
densities. Sparse galaxy samples like these will render it very hard to extract any significant signal of the nature 
of dark energy. For the exploitation of void shape statistics in such observational circumstances, significantly denser 
samples or samples over considerably large volumes will be needed. 

A second major complicating factor is the spatial bias of the halo and galaxy distribution with respect to the 
underlying dark matter field. Even when the observational data sets would consist of a densely sampled halo or 
galaxy distribution, they still may not form a fair reflection of the underlying mass distribution. 
Spatial bias is a major complication in the case of the halo population investigated in this study. 

The situation would be even more complicated for the galaxy distribution. Baryonic processes involved in the formation of galaxies 
in and around voids produce effects which are not yet fully understood. Recent work has shown that different semi-analytical 
prescriptions of galaxy formation (SAM) lead to significantly different galaxy populations in and around voids \citep{delucia06, bower06}. 
The same mass distribution may imply a biased void galaxy population in one SAM, while another implies an anti-biased population 
\citep{platenphd, platen12}. A possible solution to problems 
related to the spatial bias of haloes and/or galaxies would be to infer information from different unbiased sources. A 
promising possibility might be the use of (dark) matter maps determined by gravitational lensing. 

Another -- practical -- factor that influences our results are the instruments that we use for identifying and studying the 
void populations. The watershed void finder (WVF) succeeds beautifully in delineating the often erratic outline of voids. 
However, the DTFE density field reconstruction -- from which the WVF voids are determined -- may be quite noisy in the case 
of the sparsely sampled data encountered in the observational reality. We are in the process of investigating other 
density estimators and void finders. Preliminary results show that the MAK reconstruction formalism used by 
\citet{lavaux10} may partially alleviate this practical consideration \citep{ruwen11}.

Circumventing the complications with spatially resolving the void population with the sparse galaxy or halo 
population is the suggestion by \cite{lavaux11} to combine the signal of all sampled and observed voids via 
a scaled stacking of the voids. The resulting elongation or flattening of the stacked void may then be 
ascribed to the Alcock-Paczynski effect \citep{alcock79}. The measured size differences in radial and 
transverse direction of the stacked void can then be directly related to the product of angular diameter 
distance and Hubble parameter. This deals to a large extent with the discreteness and sparseness of the  
data sample, and \cite{lavaux11} argue and demonstrate convincingly that it leads to an highly accurate 
assessment of the dark energy equation of state \citep[also see][]{shoji12}. Nonetheless, it may still suffer 
from uncertainties on the biasing properties of the void galaxy population which, as we have seen in this 
study, may have a sizeable influence.

\section*{Acknowledgments}
\label{sec:acknowledgements}
Roman Juszkiewicz and Wojciech Hellwing strongly supported this research at the starting phase. We are grateful 
to Wojciech for many useful suggestions and fruitful discussions. We feel particularly honoured by the 
encouragement that Roman has given to this project and, now he is no longer with us, we would like to pay tribute 
to him and his many seminal contributions in our understanding of the nonlinear universe.
We also want to thank the referee, Jaan Einasto, for helpful comments. In addition, we gratefully 
acknowledge discussions with Bernard Jones, Jarno Ruwen, Marius Cautun and Johan Hidding. We thank Erwin Platen for 
discussions and for the WVF software that was used in this study. P.B. acknowledges support by the NOVA project 
10.1.3.07. K.D. acknowledges the support by the DFG Priority Programme 1177 and additional support by the 
DFG Cluster of Excellence ``Origin and Structure of the Universe''. V.P. is supported by 
the IEF Marie Curie, project `DEMO' (Dark Energy Models and Observations).


\bibliographystyle{mn2efixed}
\setlength{\bibhang}{2.0em}
\setlength\labelwidth{0.0em}
\bibliography{mn-jour,egpbib}

\appendix

\section{Models of Dark Energy}
\label{sec:modelsofDE}
In this section, we elaborate on the specific models of dark energy that we have considered in this work (the same models were used in \citet{deboni11}). Dark energy has its influence on cosmology through the Friedmann equation (equation~\ref{eqn:friedmann}). Hence, we derive $w_\mathrm{DE}(a)$ for each model, which is the only remaining missing piece from equation~\ref{eqn:friedmann}.

Our reference model is a universe containing cold DM and a cosmological constant: $\Lambda$CDM. We compare this model to four different models of time dependent dark energy. We use two quintessence models, in which the dark energy is described as a scalar field under the influence of a potential. The other two models are extended quintessence models, where the scalar field is coupled to gravity.

In the following we set $a_0 = 1$. We assume a universe with flat geometry, i.e.\ without curvature. The equations used to determine $w(a)$ are given. The resulting relations for the different models are shown in figure~\ref{fig:wvsz}.

\paragraph*{Cosmological constant}

Dark energy in a $\Lambda$CDM cosmology is modelled by a cosmological constant, or equivalently a constant $w_\Lambda = -1$ in equation~\ref{eqn:friedmann}.

\paragraph*{Quintessence}

Dark energy modelled by a scalar field $\phi$ in a potential $V(\phi)$ is called ``quintessence'' dark energy \citep{wetterich88, rp88}. This model has $w = w(a)$ and the Friedmann equation is
\begin{equation}
\begin{split}
	\left(\frac{H}{H_0} \right)^2 = \frac{\Omega_{0,m}}{a^3} + \frac{\Omega_{0,r}}{a^4} \\
	+ \Omega_{0,\phi}\exp\left( -3 \int_{a_0}^a \frac{1+w_\phi(a')}{a'}da' \right) \,,
\end{split}
\end{equation}
where
\begin{equation}
     \label{eqn:w_quintessence}
     w_\phi = \frac{P_\phi}{\rho_\phi} = \frac{\frac{1}{2}\dot{\phi}^2 - V(\phi)}{\frac{1}{2}\dot{\phi}^2 + V(\phi)} \,.
\end{equation}
Note that when the kinetic term $\dot{\phi}$ vanishes, we regain the $\Lambda$CDM value of $w=-1$. The cosmological constant can, thus, be seen as a special case of the more general quintessence model of dark energy. We can solve for $\phi$ using the Klein-Gordon equation:
\begin{equation}
	\ddot{\phi} + 3H\dot{\phi} + \frac{\partial V(\phi)}{\partial \phi} = 0 \,.
\end{equation}

The potential $V(\phi)$ determines the model's dynamical properties. We have used an inverse power law potential \citep{rp88} and a generalised inverse power law potential \citep{brax00}. The latter potential expands upon the former by including corrections from supergravity \citep{freedman76}. These models, which we will later refer to as RP and SUGRA respectively, have the following potentials:
\begin{equation}
	V_\mathrm{RP}(\phi) = \frac{\Lambda^{4+\alpha}}{\phi^\alpha}
\end{equation}
\begin{equation}
	V_\mathrm{SUGRA}(\phi) = \frac{\Lambda^{4+\alpha}}{\phi^\alpha}\exp\left( 4\pi G \phi^2 \right) \,,
\end{equation}
where $\alpha \geqslant 0$ and $\Lambda$ are free parameters. They are both tracker potentials.

\paragraph*{Extended Quintessence}
Secondly, we consider a scalar field explicitly coupled to the rest of the universal components through gravity \citep{boisseau00}. Specifically, we consider here the so-called ``extended quintessence'' (EQ) models \citep{pettorino08}. The way we represent an interaction in field theory is by adding an interaction term to the action of the field. This term is a (Lorentz invariant) product of the quantities that represent the fields that we want to interact. In our case, these are the gravitational field represented by the Ricci scalar $R$ and the EQ field $\phi$. The action then becomes \citep{baccigalupi00}
\begin{equation}
\begin{split}
S = &\int d^4x \sqrt{-g} \\
    &\left[ \frac{1}{2} F(\phi)R - \frac{1}{2}\partial^\mu\phi\partial_\mu\phi - V(\phi) + L' \right] \,,
\end{split}
\end{equation}
where $L'$ contains the terms of the Lagrangian without $\phi$. $F(\phi)$ is given by
\begin{equation}
 F(\phi) = \frac{1}{8\pi G} + \xi \left(\phi^2 - \phi_0^2\right) \,,
\end{equation}
where $\xi$ determines the strength of the interaction and $\phi_0 = \phi(t_0)$.

In what follows, we only consider the linear (Newtonian) limit. We can then approximate the effect of EQ in an N-body simulation by replacing the gravitational constant $G$ by a time dependent parameter $\tilde{G}$, given by:
\begin{equation}
	\frac{\tilde{G}}G \sim 1 - 8\pi G \xi (\phi^2 - \phi_0^2) \,.
\end{equation}
This is supported by version 3 of the GADGET code and, thus, we conveniently solve this otherwise complicated problem. Additionally, in the linear regime, the equation of state parameter $w(z)$ behaves like the normal quintessence ones. We use the RP potential for these models.

The linear approximation is valid if $w_{JBD} \gg 1$, where
\begin{equation}
\label{eqn:wJBD}
 w_{JBD} \equiv \frac{F(\phi)}{\left[\partial F(\phi)/\partial \phi\right]^2} = \frac{\frac{1}{8\pi G} + \xi \left(\phi^2 - \phi_0^2\right)}{4\xi^2\phi^2} \,.
\end{equation}
Using this relation we can determine the allowed values of $\xi$. The lower limit for $w_{JBD,0}$ (and, thus, the upper limit for the interaction term $\xi$, because $w_{JBD,0} \propto \xi^{-2}$) can be determined by observations. On cosmological scales, this limit -- as obtained using WMAP1 and 2dF data -- is set at $\left|w_{JBD,0}\right| > 120$ \citep{acquaviva05}. Because we want the interaction to be as strong as possible within observational limits we indeed set $\xi$ using this value; $w_{JBD,0} = 120$.

The two models we consider are those with a positive and a negative value of $\xi$ (referred to as EQp and EQn hence forward). These two models differ slightly in the gravitational parameter. At $z<0$, $\frac{\tilde{G}}G > 1$ for EQp and $\frac{\tilde{G}}G < 1$ for EQn. These corrections are within the few percent level.

\paragraph*{General parametrization}

We can fit equation~\ref{eqn:DEgeneralparam} to these models using a simple $\chi^2$ fitting procedure with $w_a$ as a free parameter ($w_0$ is fixed to the values that were chosen for the simulations). We find the best fits as given in table~\ref{tab:wafits} (and again in table~\ref{tab:DEparams}).

\begin{table}
\begin{tabular*}{\columnwidth}{@{\extracolsep{\fill}}lrrrrr}
\itshape Model & $\Lambda$CDM & RP & SUGRA & EQp & EQn \\ \hline
$w_a$ & $0.0$ & $0.0564$ & $0.452$ & $0.0117$ & $0.0805$ \\
$w_0$ & $-1.0$  & $-0.9$ & $-0.9$ & $-0.9$ & $-0.9$ \\
\end{tabular*}
\caption{Fits of dark energy model parameters $w_a$ to the $w(a)$ relations in figure~\ref{fig:wvsz}, determined using a $\chi^2$ fit.}
\label{tab:wafits}
\end{table}

\label{lastpage}

\end{document}